\documentclass[journal=jacsat,manuscript=article]{achemso}

\usepackage[utf8]{inputenc}
\usepackage[T1]{fontenc}

\usepackage{graphicx}
\usepackage{color}

\usepackage[version=3]{mhchem} 

\author{Sofia Zinzani}
\affiliation
{Department of Physics, University of Milan, Via Celoria 16, I-20133, Italy}
\author{Robert M. Jones}
\affiliation
{Department of Physics, King's College London, WC2R 2LS, London, UK}
\author{Mirko Vanzan}
\affiliation
{Department of Physics, University of Milan, Via Celoria 16, I-20133, Italy}
\author{Francesca Baletto}
\email{francesca.baletto@unimi.it}
\affiliation
{Department of Physics, University of Milan, Via Celoria 16, I-20133, Italy}
\alsoaffiliation
{Department of Physics, King's College London, WC2R 2LS, London, UK}
\title[An \textsf{achemso} demo]
{Morphological stability of Au-metal nanosatellites}


\begin{document}


\begin{abstract}

Hybrid metallic nanoalloys combining plasmonic and catalytic metals are essential for developing advanced photocatalysts. A promising design called core-satellites comprises a spherical nanogold dotted with smaller transition-metal clusters. While these nanoalloys' catalytic activity and hot-carriers generation have been extensively studied, their morphological stability remains poorly explored. 
Performing molecular dynamics simulations, we highlight the critical role of the transition metal in governing the morphological stability of plasmonic core-satellites.
Rh satellites exhibit the highest stability, while only 27\% Pt and 16\% Pd satellites survive after 200 ns at 600K. AuPt and AuPd quickly rearrange into single spherical nanostructures. AuPt forms icosahedra with an Au outer shell due to Au's surface diffusion. AuPd favors FCC and decahedral shapes and shows the highest Au mobility and significant Pd interdiffusion. In contrast, AuRh maintains its original shape, exhibiting a slow surface diffusion of gold onto rhodium and negligible mixing.
\end{abstract}

\begin{figure}
    \centering
    \includegraphics[width=1\linewidth]{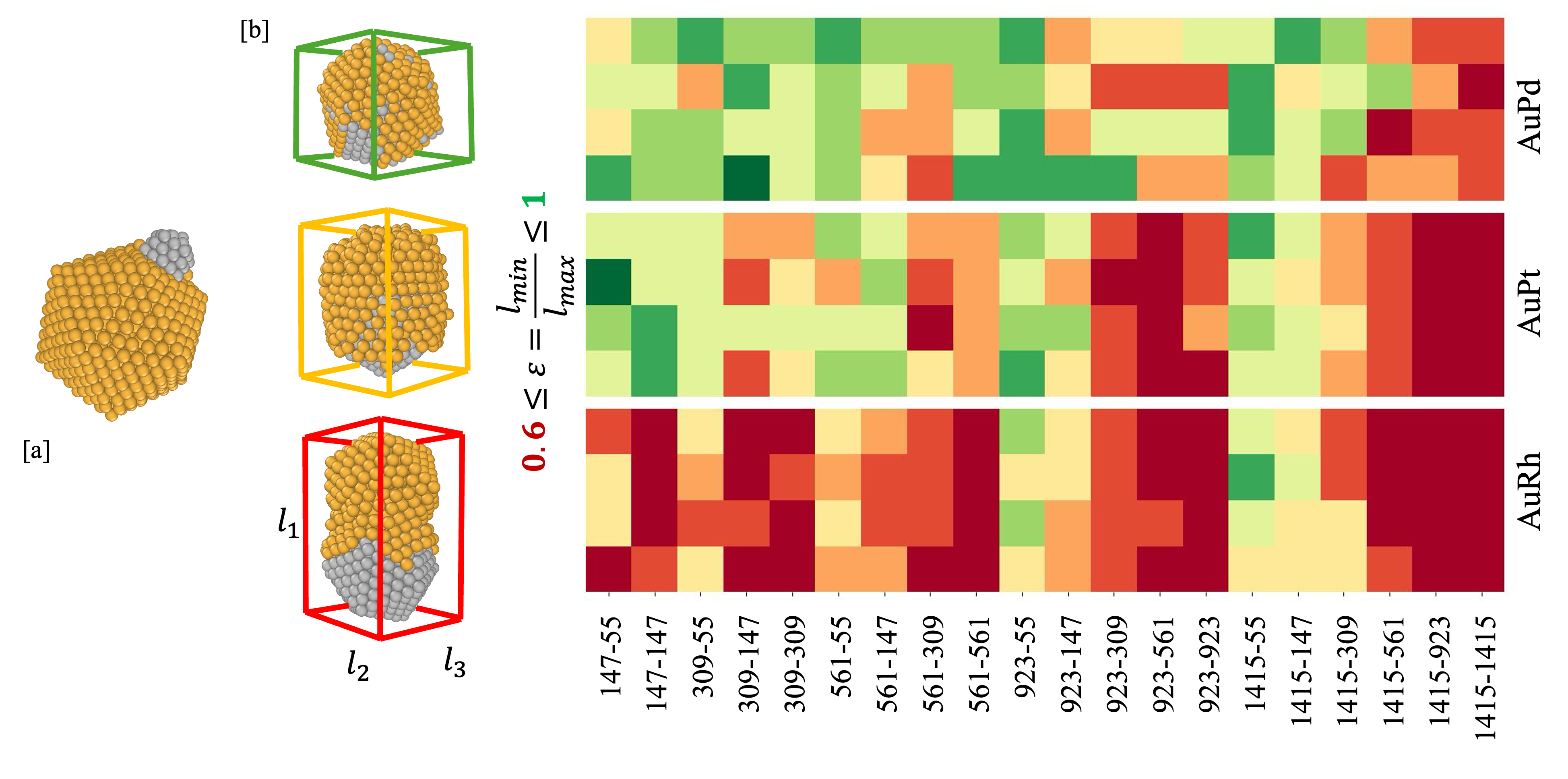}
    \caption{[a] Typical Au-CS nanoreactor with Au$_{1415}$TM$_{55}$, where gold and grey spheres refer to Au and TM atoms, respectively.
    [b] Shape asymmetry, $\epsilon$, of the considered Au-CS. $\epsilon$ is calculated as the ratio between the longest and shortest edges of the box enclosing Au-CS. $\epsilon$-heatmap versus Au-CS size -with the number of gold atoms reported first- after 200 ns: green identifies $\epsilon \rightarrow 1$ and red $\epsilon$ values close to 0.6. The top panel refers to AuPd, the middle to AuPt, and the bottom to AuRh. Each row corresponds to each independent simulation. Snapshots are visualized with OVITO.\cite{ovito}
    }
    \label{fig:eccentricity}
\end{figure}

Hybrid and Janus-like metallic nanoalloys are renowned for their exceptional photocatalytic performance, especially when a plasmonic metal, like Au, combines with a transition metal (TM). Janus Au-TM nanostructures enable a synergistic interaction between strong optical capabilities and catalytic properties of their components, paving the way to plasmocatalysts, sensing, spectroscopy, and photothermal therapy.\cite{Ezendam2022,hassan2023au-based-461,D1TA03646G, Liu2024} \\
In particular, hybrid Au-nanoalloys featuring a large gold core adorned asymmetrically with smaller TM-satellites (Pt, Pd, Rh) are shown to be efficient plasmonic photocatalysts.\cite{Zhang2017, Liu-Corma2023, salmn-gamboa2020rational-b52, herran2022tailoring-b34}
The rational behind the core/satellite design is that they can be seen as antenna-reactor complexes, where the plasmonic core (Au) acts as the antenna and the catalytic satellite (TM) serves as the reactor.\cite{Yang2024, Yuan2022,li2017balancing-9c2} Once the antenna absorbs the external electromagnetic reaction, the induced local field perturbs the electronic structure of the satellite/reactor, promoting the generation of hot-carriers, enhancing the local electric field, and increasing the local temperature.\cite{Vanzan2024, Jiang2023} 
Gold-based nanoparticles are especially well-suited for the antenna role due to their strong Localized Surface Plasmon Resonance (LSPR) in the visible range and their resistance to oxidation.
When excited by external radiation at their LSPR, AuRh, AuPt, and AuPd nanoalloys offer a remarkable photocatalytic activity towards critically important reactions such as hydrogen and oxygen evolution, carbon dioxide reduction and small organic molecule oxidation reactions.\cite{C2CP23974D,rodrigues2021,Peneau.PCCP.2013,Al-Rifai2017,RODRIGUES2022140439} 
Since the pioneering synthesis by Yoon et al. \cite{Yoon2012},  engineering routes are constantly proposed to form Janus Au-core/TM-satellites nanoreactors (Au-CS).\cite{Fan2025,Trinh2022,Zhao2020, coviello2022}
The morphological stability of Au-CS is critical to determine their plasmocatalytic performance, as size, shape, and the chemical ordering inherently affect their optical and catalytic properties.
\cite{Lischner2023}\\
Understanding how chemical species distribute over time in multicomponent materials is essential for controlling their functional properties. At the nanoscale, surface diffusion, surface segregation and miscibility can significantly differ from their bulk counterpart.\cite{Yang2020,Eom2021,Parsina2010,Basso2025,Loevlie2023,Ter-Oganessian2024} For instance, AuRh are immiscible in the bulk but miscible at the nanoscale.\cite{Chen2024} \\
{\em A-priori}, following Hume-Rothery's rules and surface energy considerations, one expects that Au will tend to move at the surface on a timescale that depends on the likelihood of Au-adatoms to surface diffuse and the TM to interdiffusion.
However, an Au-shell on a TM-core is detrimental to the photocatalytic applications of Au-nanoalloys as it reduces the number of surface active sites and eventually compromises the optical properties of the antenna.\cite{salmn-gamboa2018optimizing-00e}
Therefore, morphological information regarding the structural and chemical ordering of Janus Au-CS at finite temperatures is highly needed.\cite{Li2020} \\
Our primary interest is determining whether the core/satellite morphology of an Au-nanoreactor is preserved during finite temperature dynamics through molecular dynamics.
Our work benefits from other numerical studies on the sintering of metallic nanoparticles. However, they mainly focus on the methodology and exploring deposition, orientational and environmental effects and not on investigating the stability of Au-CS,\cite{Cioni2024,Itina2024, Grammatikopoulos2023, Nelli31122023, FerrandoNanoalloys, Zinzani2024} with AuRh poorly studied.\cite{Vanzan_aurh, Piccolo2016} Here, we concentrate on the Janus-stability of Au-TM and the TMs' miscibility over time. \\
We run sizes of the Au-core and the TM-satellite between 55 and 1415 atoms (0.6-4.6 nm), considering the Au-rich regime so that the TM concentrations are between 3.7 and 50\%. Among the three TMs, AuRh is the most promising system for forming a stable nanoreactor with much higher stability than AuPt and AuPd. 
We use home-made software freely available\cite{Lodis} to investigate the evolution of Au-CS, starting from a simple geometrical model obtained by soft-landing of a smaller TM icosahedron (Ih) onto an Au-core, still displaying an Ih shape.
We limit our investigation to icosahedral seeds, the most reminiscent of a sphere. The Au-CS starting configuration is a double-Ih (dIh), as depicted in Figure \ref{fig:eccentricity}a. 
After a sintering time of 1 ns, we propose an in-depth structural and chemical analysis of 200 ns long trajectories at 600 K to elucidate the underlying mechanisms driving the dynamical morphological rearrangements. We employ a timestep of 5 fs and an Andersen thermostat to keep the temperature constant. We accumulate four independent simulations per system.
The interparticle interaction is derived within the second-moment approximation of the tight-binding theory,\cite{rosato1989thermodynamical} with the parameters used in Table S1-S3 in the Supporting Information (SI).

First, we discuss the persistence of a structural asymmetry. We quantify it from the factor $\epsilon$ as reported in Figure \ref{fig:eccentricity}. AuRh exhibits long-lasting asymmetrical shapes, as depicted in Figure \ref{fig:eccentricity}b. 85\% of AuRh are still far from a spherical shape after 200 ns. 
Exceptions appear for Rh concentrations $\leq 14\%$ and an Au-seed of at least 923 atoms. In such cases, the asymmetry factor $\epsilon$ approaches one. \\
In contrast, AuPd and AuPt CS nanostructures exhibit a higher tendency to adopt spherical shapes. 
AuPd CS evolve into spherical morphologies in about 80\% of cases. Notable exceptions occur in larger systems, e.g., an Au-core$_{1415}$ and a Pd-satellite $\geq 561$ and for the Au$_{561}$Pd$_{309}$. 
55\% of AuPt CS rearrange into spherical shapes after 200 ns in various compositions, although asymmetrical shapes occur for the Au core with more than 309 atoms depending on the concentration of TM.

\begin{figure}
    \centering
    \includegraphics[width=1\linewidth]{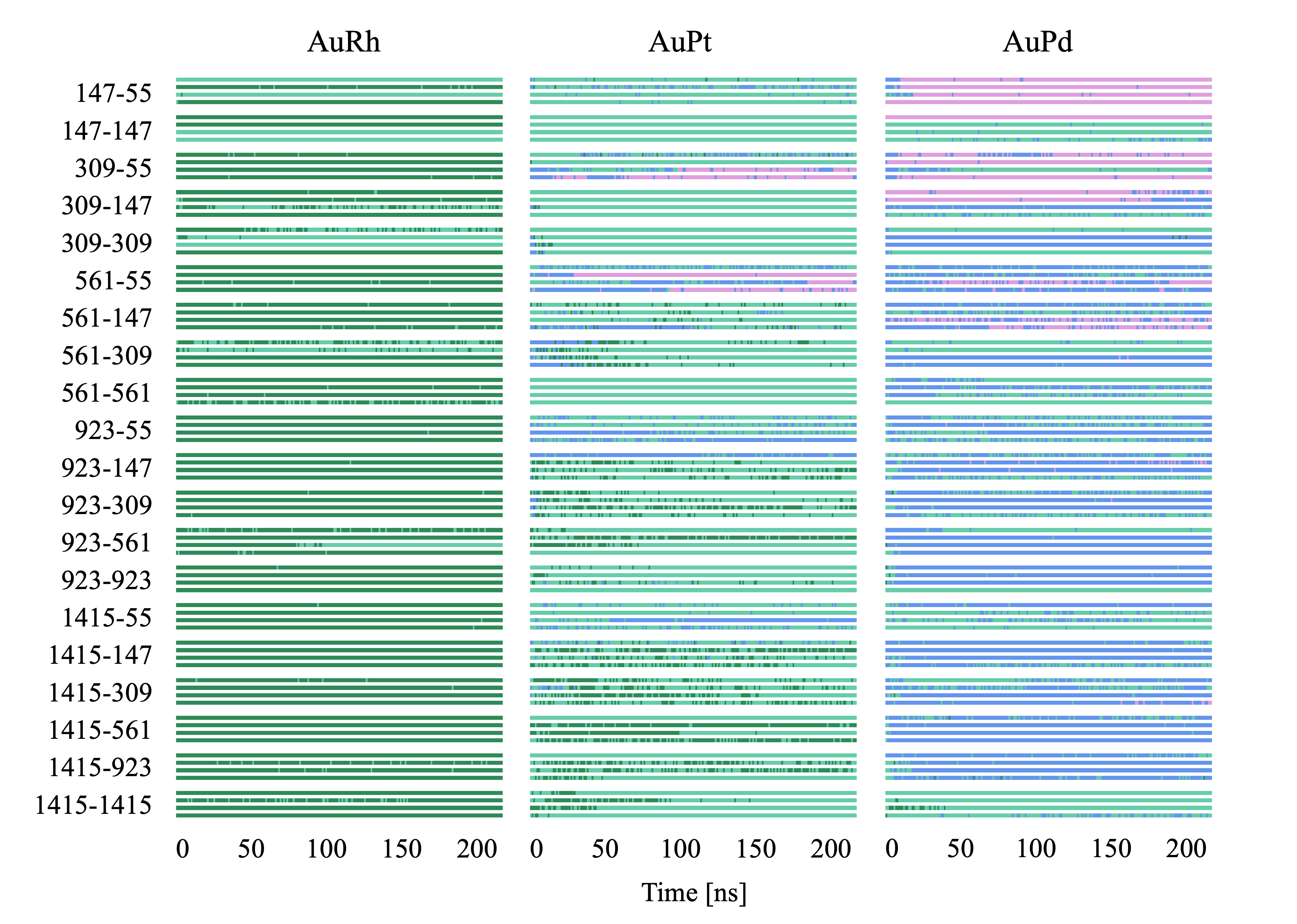}
    \caption{Structural evolution of Au-CS. From left to right, AuRh, AuPt, AuPd. Top to bottom is the Au-CS size, with the number of atoms in the Au-core reported first. Each line per size corresponds to an independent simulation. Each trajectory is sampled every 1 ns. The structural identification is based on the CNA as described in the text and in SI. dIh is in dark green, Ih is in light green, Dh is blue, and FCC is plum, respectively. }
    \label{fig:shape}
\end{figure}

Figure \ref{fig:shape} illustrates the structural dynamics of the three nanoalloys. We analyze the structural evolution of Au-CSs combining the Common Neighbors Analysis (CNA) and their Patterns (CNAP).\cite{art:sapphire, Baletto2019, Nelli31122023} More details on this analysis are available in the SI. \\ 
We classify the shapes of the nanoalloys into the following categories: dIh, Ih, decahedra (Dh), and FCC, even when defects are present. Notably, none of the sizes examined corresponds to a magic number representing geometrical shell closure. Each Au-CS dynamics proceed without undergoing phase transitions and the morphological rearrangements occur through solid-solid transformations, in contrast to other bimetallic nanosystems.\cite{Rossi2018, C7CP01397C} \\
AuPd CSs tend to evolve into Dh 65\% of the cases, while FCC structures appear only in a few selected small sizes, with very few instances of Ih formation. None of the AuPd nanostructures preserves a dIh shape, indicating the low thermal stability of AuPd CS.
Although Au$_{309}$Pd$_{55}$, Au$_{309}$Pd$_{147}$, Au$_{561}$Pd$_{55}$ and Au$_{561}$Pd$_{147}$ primarily exhibit an FCC arrangement, they present rapid oscillations among the Dh, FCC, and Ih motifs, reflecting their low thermal stability. \\
In the case of AuPt, there is size-dependent competition among structural motifs. For smaller sizes, those below Au$_{923}$Pt$_{147}$, we do not observe the nanoalloy preserving the initial dIh shape; instead, we witness significant competition between Ih and Dh, with only a few FCC structures, mainly when the initial satellite is a Pt$_{55}$. At larger sizes, AuPt discloses a dIh $\rightarrow$ Ih transition. \\
AuRh CS exhibit the highest resistance at 600 K. At larger sizes, the dIh shape lasts, except for Au$_{923}$Rh$_{561}$ that transforms into an Ih. For smaller sizes, specifically when the Au seed contains less than 561 atoms, the AuRh system rearranges into an Ih configuration. In these cases, the Ih center is preferentially localized within the Rh satellite and originates from the same atom.

\begin{figure}
    \centering
    \includegraphics[width=1\linewidth]{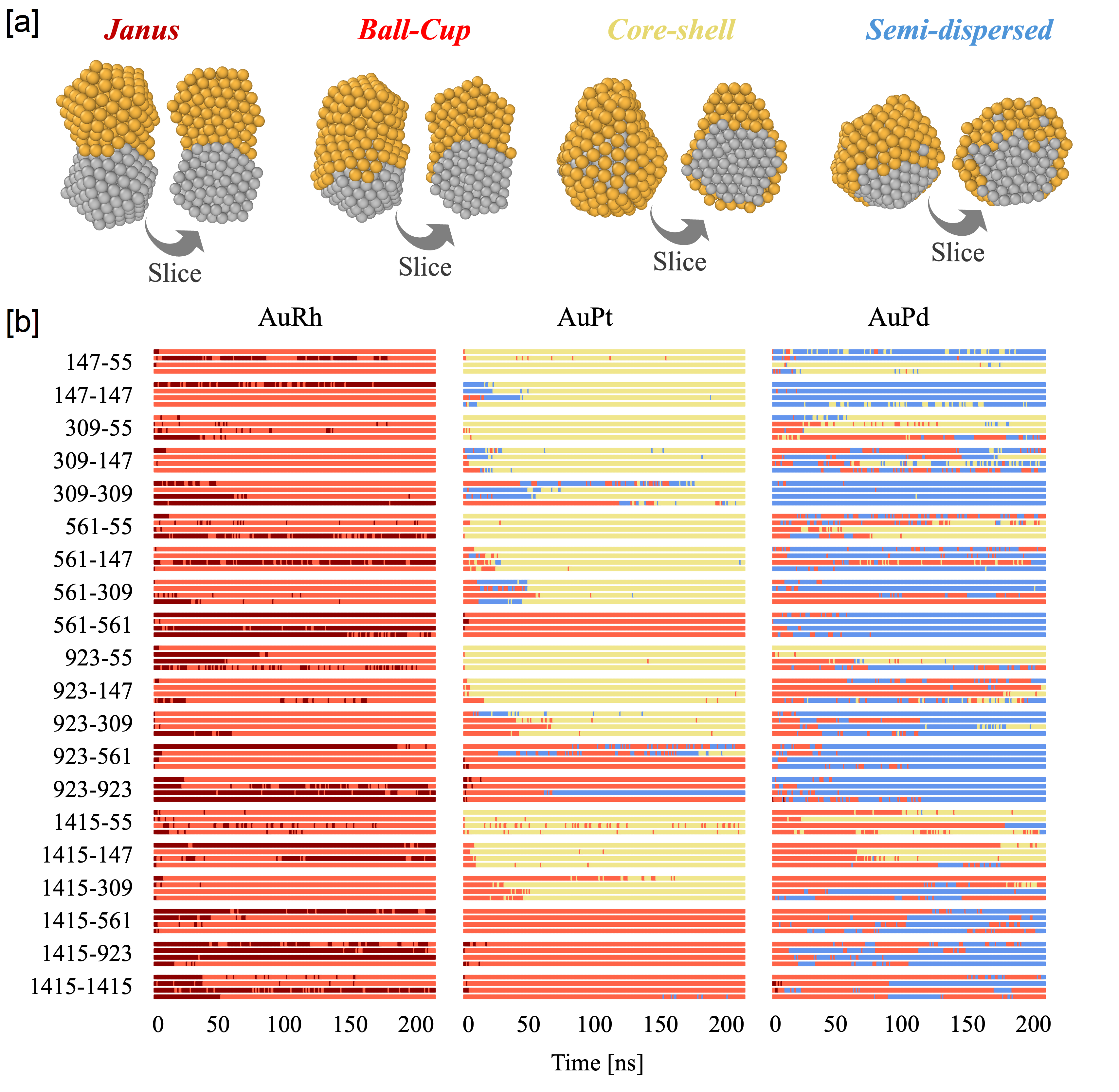}
    \caption{[a] nomenclature of the four chemical ordering: Janus, ball-cup (BC), core-shell (S@C), and semi-dispersed (SD). Gold and grey spheres refer to Au and TM atoms, respectively. Representation sliced along the CS-axes are included in order to show the inner part of AuCS.
    [b] chemical ordering evolution of Au-TM CS. From left to right, AuRh, AuPt, AuPd. Top to bottom is the Au-CS size, with the number of atoms in the Au-core reported first. Each line per size corresponds to an independent simulation; trajectory sampled every 1 ns Each point represents the chemical ordering classification of the structure. Dark red, light red, khaki and light blue refer to Janus, BC, S@C and SD, respectively.
    }
    \label{fig:Chem.ord}
\end{figure}

\begin{figure}
    \centering
    \includegraphics[width=1\linewidth]{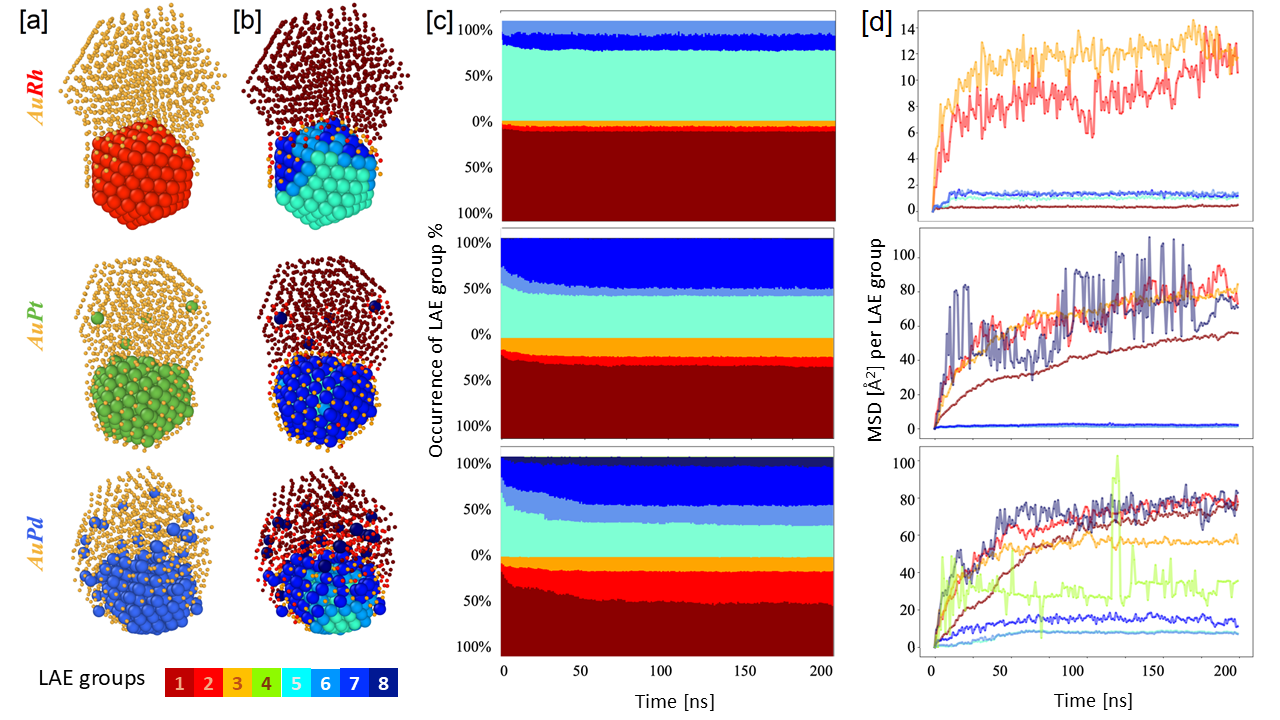}
    \caption{Au$_{923}$-TM$_{309}$ examples. [a] Typical snapshots of AuRh (top), AuPt (middle), and AuPd (bottom), taken at 200 ns. Smaller gold spheres represent Au atoms while larger spheres indicate TM atoms. Rh, Pt, and Pd depicted in red, green, and blue, respectively. [b] same snapshot as in [a] where atoms are colored based on their LAE group. 1 to 4 for Au atoms and 5 to 8 for the TM as in the legend. [b] Atomic occurrence, with respect to the initial seed, in each LAE group over time. [c] MSD analysis per each LAE group. Colors in [b,c,d] refer to legend on the bottom of panel [a, b].}
    \label{fig:MSD}
\end{figure}

To understand how different chemical elements are distributed within a nanoalloy as a function of time, we combine a chemical layer-by-layer (cLBL)\cite{Zinzani2024} and local atomic environment (LAE)\cite{art:sapphire} analysis. 
CLBL calculates the occurrence of Au (TM) atoms in every layer, sectioning each nanostructure perpendicularly to the axis connecting the centre-of-mass of the two metallic seeds, called the CS-axis. 
The LAE, instead, describes the chemical network around each atom, estimating the number of homo- and hetero-bonds within a fixed cutoff between the first and second shells of coordination. LAE splits atoms into eight groups, four per element: (LAE$_1$) Au (TM, LAE$_5$) with only homo-bonds; (LAE$_2$) Au (TM, LAE$_6$) with less than three hetero-bonds; (LAE$_3$) Au (TM, LAE$_7$) with more than nine hetero-bonds identifying encapsulated or dispersed atom within the other metal; (LAE$_4$ for Au and LAE$_8$ for TM) otherwise, atoms that form the interface between the two metals. The analysis of all systems are available in the SI, see Figures S7-S8-S9. \\
We distinguish five different chemical orderings, depicted in Figure \ref{fig:Chem.ord}a, depending on the relative occurrence of the two metals in each layer and by the presence of certain LAE: Janus, ball-cup (BC), shell@core (S@C), semi-dispersed (SD, sometimes named doping \cite{Chen2021}), fully mixed (not shown in the picture). \\
BC is when Au only partially covers the TM leaving several layers of the initial TM seed exposed.\cite{C6CY01107A}
Au-CS nanostructures correspond to both Janus and BC chemical distribution.
S@C, one of the most popular elemental distributions, can be seen as the evolution of a BC when one of the metals, here Au, forms an outer shell, meaning its occurrence is more than 75\% in the top and bottom layers with respect to the CS-axis. In a vacuum, it is unlikely to have an inverse S@C, with the shell made of the TM, and it is not considered further. 
SD is the BC evolution alternative to the S@C, when LAE$_8$ is not zero. It refers to situations where isolated TM atoms are dispersed in the Au seed. 
The fully mixed, which can be either random or organized, is when the segregation is negligible. As we never observed it on the 200 ns timescale, it is not considered further. \\
After equilibration at 600K, the initial Janus ordering of the Au-CS could (i) remain unchanged, (ii) begin interdiffusion at the Au-TM interface, or (iii) evolve into a BC. 74\% of AuRh nanostructures exhibit a Janus ordering while the remaining 26\% are BC. That percentage is still as high as 93\% for AuPd -of which less than 10\% are Janus- and drops to 84\% for AuPt -where 19\% are Janus and the remaining are BC. 
Direct transitions from a Janus to a mixed interface are never observed as Au diffuses onto the TM seed before TM interdiffusion can start.
During the dynamical evolution, the scenario substantially changes, as highlighted in Figure \ref{fig:Chem.ord}.
After 200 ns, most AuRh nanosatellites keep a BC (84\%) or still have a Janus (16\%) chemical distribution. S@C or SD configurations are never observed, in accordance with experimental results.\cite{Chen2021, Chen2024} On the other hand, Janus ordering is absent in both AuPd and AuPt. Segregated configurations are sparsely found for AuPt (28\%) and even less for AuPd (16\%). In any event, no ordered alloys are formed, and the dissolution of the whole TM within the Au seed is rare, as expected for these kind of systems.\cite{Chatzidakis2017,Chen2021,Hwang2024} \\
Interestingly, comparing Figure \ref{fig:shape} and Figure \ref{fig:Chem.ord}, there are no clear correlations between the structural and chemical rearrangements, suggesting that transformations in chemical architecture can proceed without corresponding geometrical changes, and vice versa. For example, besides AuPt and AuPd show significant morphological changes, their elemental redistribution follows two different mechanisms, as detailed below.
While solid-solid transition mechanisms are detailed elsewhere\cite{Baletto2019,Rossi2018}, to comprehend the kinetics and the origin of Au-CS chemical distribution rearrangements, we calculate the mean square displacement (MSD) for each LAE-group. Figure \ref{fig:MSD} depicts the mobility of Au and TM LAE-groups in paradigmatic examples. A complete overview of all the considered systems and their MSD are shown in SI (see Figure S7-S8-S9). 

In all systems, Au is more mobile than TM, in agreement with adatom diffusion on (111) surfaces\cite{Mavrikakis2017_diffusion}, except for a quarter of cases for Au$_{923}$Rh$_{55}$ and Au$_{1415}$Rh$_{309}$, see Figures S7-S8-S9 in SI. \\
We observe comparable mean mobility of Au atoms in for AuPd and AuPt, averaging over time and for all the four LAE groups of Au we have about 50 \AA$^2$. In AuRh, the same averaged MSD falls below 15 \AA$^2$; see Figure S10 in SI.\\
Maintaining a CS structure for AuPt and AuPd is challenging. 
AuPt tends to evolve through the sequence Janus $\rightarrow$ BC $\rightarrow$ CS,  with 70\% of cases leading to a S@C. Such elemental evolution mechanism is supported by the high mobility of Au atoms with,  typical MSD values for LAE$_3$ $\geq$ those of LAE$_2 >$  LAE$_1$ , see Figure S8. There is a size effect, as with closer MSD values for Au-seed less than 923. The high mobility of Au in LAE$_2$ and LAE$_3$ is favored by the strong Ih-character and the the structural solid-solid transitions observed in Figure \ref{fig:shape}. The number of Au dispersed in TM is negligible.\cite{Chatzidakis2017,Pedrazo-Marzan2022}   
The BC ordering is kept only at a few sizes, namely Au$_{561}$Pt$_{561}$, Au$_{923}$Pt$_{561}$, Au$_{923}$Pt$_{923}$, and combinations of Au$_{1415}$Pt$_{561}$, Au$_{1415}$Pt$_{923}$ and Au$_{1415}$Pt$_{1415}$, when MSD of LAE$_2$ oscillates considerably and the Au-atoms of LAE$_1$ are less mobile. Notably, the transition BC to S@C may involve the formation of metastable SD configurations, and the lifetime of these intermediate states appears to be independent of both size and composition. Only one simulation remains trapped in an SD configuration after 200 ns, but it is statistically irrelevant. \\
The averaged mobility of Pt atoms is just less than 25 \AA$^2$. Notable deviations are the Au$_{561}$Pt$_{561}$, Au$_{923}$Pt$_{923}$, Au$_{1415}$Pt$_{561}$, systems and half of the cases at Au$_{561}$Pt$_{55}$, where the averaged mobility of Pt atoms drops almost to zero, see Figure S10. We also observe a considerably high proportion of Pt with an intermediate number of hetero bonds, see Figure \ref{fig:MSD}, LAE$_7$ (blue), suggesting that Pt does not dissolve inside Au but prefers to diffuse into subsurface sites. \\ 
AuPd evolves from a Janus to a BC ordering and in approximately 60\% of cases, it develops an SD ordering. After a BC ordering, AuPd shows an S@C configuration a quarter of the times, and the other cases, it remains trapped in a BC (15\%). This finding aligns with the tendency of AuPd to form a mixed alloy with an Au-rich outer shell.\cite{Neyman2021} Generally, the Pd MSD averages around 10 \AA$^2$ for small Au seeds (147, 309 and still Au$_{561}$Pd$_{55}$)  while at larger sizes it apporaches 20 \AA$^2$. \\
We note that AuPd shows several oscillations among different chemical orderings. However, when they occur, their residency time varies between independent simulations without any recognizable pattern. The tendency to oscillate among chemical distributions and to undergo structural transformations towards Dh and FCC (Figure \ref{fig:shape}) is motivated by the high mobility of Au atoms in an Au-rich environment (LAE groups 1 and 2, dark and light red, respectively) that is approximately equivalent to the mobility of Au atoms at the interface with Pd (LAE$_3$, yellow). See, for example Figure S9.
Moreover, we observe a high mobility of Pd atoms diffusing inside the Au (LAE$_8$, dark blue) and even the motion of the few dispersed Au atoms in the Pd (LAE$_4$, light green). Taking into account our previous study on AuPd,\cite{Zinzani2024} the survival of AuPd CS might be strength if they are embedded in a Pd-strongly interacting environment. \\
Both Figure \ref{fig:shape} and Figure \ref{fig:Chem.ord} show the strong tendency of AuRh to remain highly segregated and to mantain a dIh or an Ih shape. The only mobile Au atoms are those at the interface (LAE$_4$), followed by Au with few hetero bonds (Figure \ref{fig:MSD}, LAE group 2 in red) followed by Au atoms at the interface. 
In any event, the mobility of Au atoms is a quarter of that with Pd and Pt. The mobility of dispersed Au atoms is not negligible, although their percentage is less than 1\%. The mobility of Rh atoms in any LAE group is less than 1\AA$^2$ across all the considered cases. 
The tendency of AuRh to keep a highly segregating pattern reflects the strong immiscibility of the two metals and  supports the potential of this system for nano-antenna applications. Such tendency is due to the low mobility of Au and Rh atoms, except at small sizes and for high Au concentration, in agreement with experimental results.\cite{Yang2024} \\
In summary, we model the morphological stability of Au-TM core-satellite structures at finite temperature over a timescale significantly exceeding that of hot-carrier decay. The initial Au-TM core-satellite configuration is schematically represented as Au icosahedral seeds adorned with a smaller TM-icosahedron. Our study investigates the morphological evolution of AuRh, AuPt, and AuPd within a size range of 0.6 to 7 nm. Our findings reveal a complex interplay among size, composition, surface diffusion, and interdiffusion, with distinct system-specific trends emerging across multiple characteristic parameters. \\
AuPt and AuPd exhibit extensive morphological transformations without undergoing phase transitions, a consequence of the high mobility of Au atoms on these transition metals. Additionally, Pd demonstrates a pronounced tendency for interdiffusion within the Au matrix. In contrast, AuRh emerges as the most stable configuration for long-lived nanosatellites, reinforcing its viability for nanoreactor applications. The observed trends offer fundamental insights into the rational design of nanoalloys for specific catalytic and plasmonic applications.

\section{Author Information}
Corresponding Author \\
Francesca Baletto - Department of Physics, University of Milan, Milan, 20133, Italy - ORCID: 0000-0003-1650-0010; email: francesca.baletto@unimi.it \\
Authors \\
Sofia Zinzani - Department of Physics, University of Milan, Milan, 20133, Italy - ORCID: 0009-0009-6961-7766  \\
Robert M. Jones - Department of Physics, King's College London, WC2R 2LS, London, UK - ORCID: 0000-0002-5422-3088 \\
Mirko Vanzan - Department of Physics, University of Milan, Milan, 20133, Italy - ORCID: 0000-0003-3521-8045 \\

\section{Author contributions}
Starting from an original idea of FB and RMJ, SZ, MV, and RMJ perform the coalescence simulations. SZ implement the analysis code starting from a software by RMJ.\cite{art:sapphire} SZ analyzes and characterizes all the trajectories including their structural and chemical classification. All authors contributed to the manuscript's writing. 

\section{Conflicts of interest} 
There are no conflicts to declare.

\section*{Data availability}
The analysis tools are available on the group GitHub page \texttt{github.com/nanoMLMS}.
The molecular dynamics trajectories are stored and publicly available at Unimi Dataverse after publication.

\section{Acknowledgements}
SZ acknowledges the University of Milan and ISC SRL for financially supporting her PhD studentship (D.M. 118/2023 PNRR).  MV thanks the University of Milan for financial support for his "Assegno A" on "The beauty of nanoparticles". RMJ acknowledges Royal Society funding under URF\textbackslash R1\textbackslash 231460. FB and RMJ acknowledge the EPSRC (EP/W017075/1). FB thanks MONSTER, a project of the Italian National Centre for HPC, Big Data and Quantum computing (ICSC, CUP B93C22000620006) funded by the European Union - NextGenerationEU, through PNNR. All authors benefit of INDACO, a High-Performance Computing platform at the Università degli Studi di Milano, for computational resources. 
FB and RMJ thank the constructive discussion with prof. A. Zayats (KCL).
\begin{suppinfo}

The Supporting Information is available free of charge at XXX \\
Second-moment approximation of the tight-binding model (SMA-TB) potential parameters; Threshold selection for first-neighbors identification; CNA and CNAP thresholds for shape identification; Chemical Ordering classification criteria; LAE groups and MSD details.
\end{suppinfo}


\providecommand{\latin}[1]{#1}
\makeatletter
\providecommand{\doi}
  {\begingroup\let\do\@makeother\dospecials
  \catcode`\{=1 \catcode`\}=2 \doi@aux}
\providecommand{\doi@aux}[1]{\endgroup\texttt{#1}}
\makeatother
\providecommand*\mcitethebibliography{\thebibliography}
\csname @ifundefined\endcsname{endmcitethebibliography}  {\let\endmcitethebibliography\endthebibliography}{}

\end{document}


\section{SMA-TB potential parameters}
The Hamiltonian of our system is given by the metal–metal interaction: $E_i^{M-M}$ per each atom $i$ 
\begin{equation}
   E_{tot} = \sum_{i} E_i = \sum_{i} E_i^{M-M} \mbox{~.}
\end{equation}
The metal-metal interaction is modelled according to the second moment approximation (SMA) in the tight-binding (TB) model \cite{art:potrosato}, which is widely used to mimic the growth of NPs and NAs \cite{Baletto2000PRL, Mariscal2005, Baletto2002JPC}.
In such a framework, the total energy of each atom $i$ contains an attractive many-body term and a 2-body repulsive Born-Mayer term as
$$
E_i^{M-M}=\sum_{j \neq i}^{n_v} A_{\alpha, \beta} e^{-p_{\alpha, \beta}\left(\frac{r_{i j}}{r_{\alpha, \beta}^0}- 1\right)}-\sqrt{\sum_{j \neq i}^{n_v} \xi_{\alpha, \beta}^2 e^{-2 q_{\alpha, \beta}{\left(\frac{r_{i j}}{r_{\alpha, \beta}^0}-1\right)}}} \mbox{~~~,}
$$
where $r_{ij}$ is the atomic pair distances; $r_{\alpha, \beta}^0$ is the reference lattice parameter of individual chemical species $\alpha, \beta$, or their arithmetic mean for hetero-pairs; $n_v$ is the number of neighbours within a radius between the bulk second and third neighbour distances.
The parameters $A_{\alpha, \beta}$, $\xi_{\alpha, \beta}$, $p_{\alpha, \beta}$, and $q_{\alpha, \beta}$, see Tables \ref{T:RGL_Pd}, \ref{T:RGL_Pt}, \ref{T:RGL_Rh} are fitted to reproduce the experimental values of the cohesive energy, elastic constant, lattice parameter and bulk modulus. 
\begin{table}[hp]
    \centering
\caption{TB-SMA parameters used for AuPd nanoalloys as from Ref. \cite{Borbon2013, Fortunelli2008}}
    \begin{tabular}{ l|c|c|c|c } 
    \hline
   Chemical & $p$ & $q$ & $A$ [eV] & $\xi$ [eV]\\
    \hline
    Au-Au & 10.139 & 4.033 & 0.2095& 1.8153\\ 
    \hline
    Pd-Pd & 11.0 & 3.794 & 0.1715 & 1.7019 \\
    \hline
    Au-Pd &10.543 & 3.8862 & 0.1897 & 1.7536
    \label{T:RGL_Pd}
    \end{tabular}
\end{table}
\begin{table}[hp]
    \centering
    \caption{TB-SMA parameters used for AuPt nanoalloys as from Ref. \cite{dessens-felix2013structural-312, Borbon2013}}
    \begin{tabular}{ l|c|c|c|c } 
    \hline
    Chemical & $p$ & $q$ & $A$ [eV] & $\xi$ [eV]\\
    \hline
    Au-Au & 10.229 & 4.036 & 0.2061 & 1.790\\ 
    \hline
    Pt-Pt & 10.612 & 4.0004 & 0.2975 & 2.695 \\
    \hline
    Au-Pt & 10.42 & 4.02 & 0.25 & 2.20 \\
    \hline
    \end{tabular}
    \label{T:RGL_Pt}
\end{table}
\begin{table}[hp]
    \centering
    \caption{TB-SMA parameters used for AuRh nanoalloys as from Ref. \cite{Rossi2017}}
    \begin{tabular}{ l|c|c|c|c } 
    \hline
    Chemical & $p$ & $q$ & $A$ [eV] & $\xi$ [eV]\\
    \hline
    Au-Au & 12.500 & 3.550 & 0.1291 & 1.5222\\ 
    \hline
    Rh-Rh & 18.450 & 1.867 & 0.0629 & 1.6602 \\
    \hline
    Au-Rh & 9.529 & 1.7893 & 0.1203 & 1.5160 \\
    \hline
    \end{tabular}
    \label{T:RGL_Rh}
\end{table}
\section{Threshold selection for first-neighbors identification}
\begin{figure}[h]
    \centering
    \includegraphics[width=0.5\linewidth]{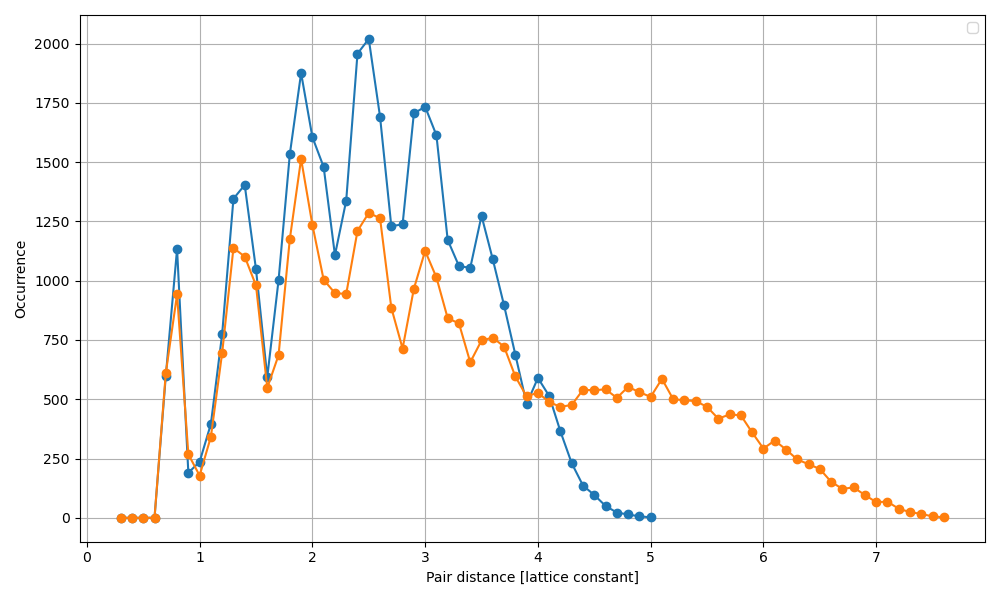}
    \caption{Global PDDF for a $Au_{147}-Pt_{55}$ atoms after the equilibration time (1 ns) (in orange) and after 200ns (in blue)}
    \label{fig:PDDF_Pt}
\end{figure}

\begin{figure}[h]
    \centering
    \includegraphics[width=0.5\linewidth]{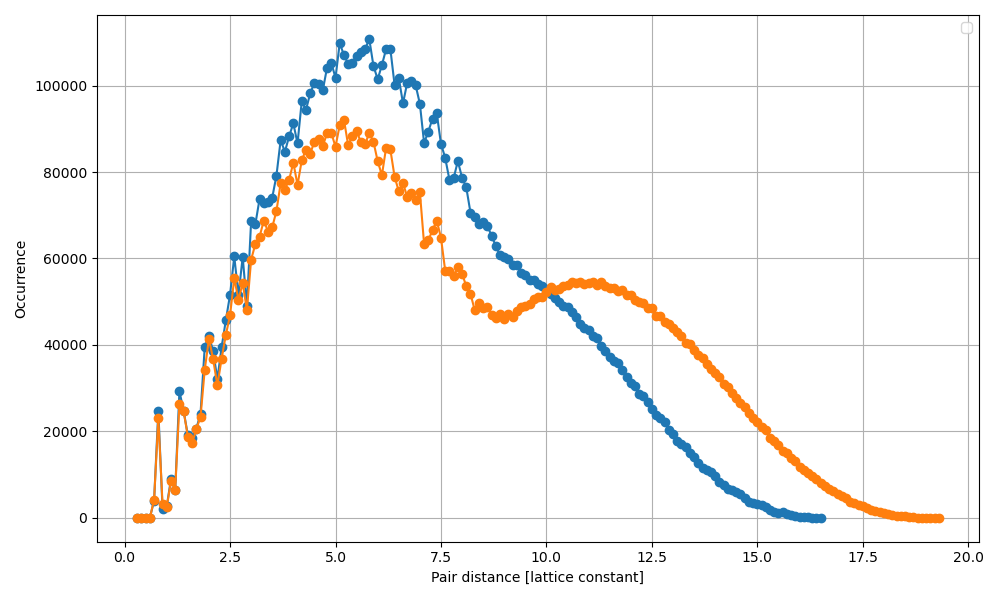}
    \caption{Global PDDF for a $Au_{1415}-Rh_{1415}$ atoms after the equilibration time (1 ns) and after 200ns (in blue).}
    \label{fig:PDDF_Rh}
\end{figure}

The distribution of pair-atomic distances (pair-distance distribution function, PDDF) is a crucial quantity to characterize the geometry and chemical ordering of a nanostructures.
We define the \textit{ pair distance} as the distance $d_{i j}$ between atoms $i$ and $j$. Given the Cartesian coordinates of the atom $i$ and $j$, $(x_i,y_i,z_i)$ and $(x_j,y_j,z_j)$ respectively,
\begin{equation}
d_{i j}=\sqrt{\left(x_i-x_j\right)^2+\left(y_i-y_j\right)^2+\left(z_i-z_j\right)^2} \mbox{~~~.}
\end{equation}
From the definition of the pair distance, its distribution function can also be studied.
The first PDDF minimum is useful to determine the cut-off distance to label the nearest neighbors. This enables the definition of an adjacency matrix and hence of all coordination properties, including common-neighbor analysis. The first neiboughs are defined as the atoms falling within a distance of 0.85 times the metal lattice parameter from the reference atom.
To clarify this argument the pair distance distribution functions of $Au_{147}-Pt_{55}$ and $Au_{1415}-Rh_{1415}$ nanoalloys, at the initial time of our simulation and after 200 ns are reported in Figures \ref{fig:PDDF_Pt}, \ref{fig:PDDF_Rh}.

\section{CNA and CNAP thresholds for shape identification}
\begin{figure}[H]
    \centering
    \includegraphics[width=.5\linewidth]{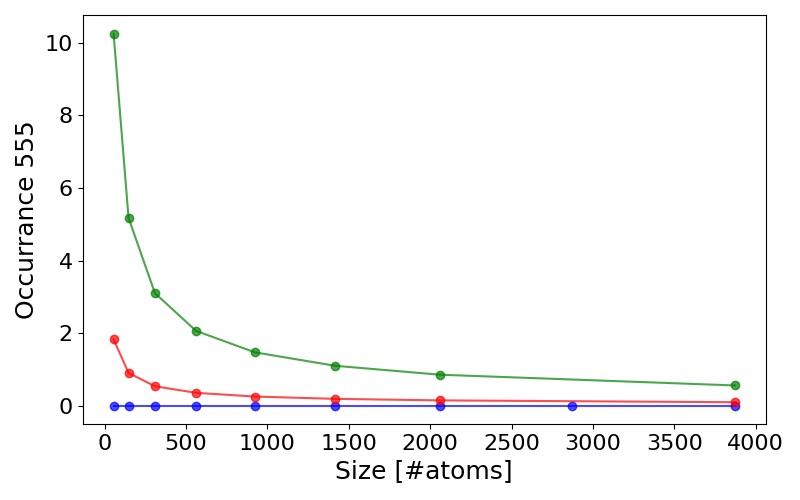}
    \caption{CNA (555) reference used to define the classification criteria based on pure structural motifs—icosahedral (Ih), Ino decahedral (InoDh), and cuboctahedral (Co)—represented in green, blue, and red, respectively. The plot shows their occurrence as a function of particle size in monometallic systems, illustrating the size range over which each structural type predominates.}
    \label{fig:cna_555_references}
\end{figure}

To classify the geometrical environments of core and surface atoms, we can introduce a characterization procedures: common neighbour analysis (CNA).
CNA characterizes pair of nearest neighbors depending on the local connectivity of their shared neighbors. CNA signatures are of the form (r,s,t) such that r is the number of nearest-neighbours common to both atoms in the pair; s is the number of bonds between shared neighbours and t is the longest chain which can be made from bonding atoms if they are nearest neighbours.
While an individual CNA signature describes the local environment of a pair of neighbours, the estimate of the percentage of how many pairs contributes to selected CNA signatures provides information on the overall nanoparticle’s structure, allowing a fast classification into expected geometrical family, as icosahedral, decahedral, or FCC-like.
In Figure \ref{fig:cna_555_references}, we present the (555)
signatures for the ideal shapes of icosahedral (Ih), Ino decahedral (InoDh), and cuboctahedral (Co) nanoparticles within a size range that encompasses the one used in our simulations. By connecting the points, we can define the "threshold" areas that were used to identify the shapes of the systems, depending on their size.
In our analysis the classification was performed by combining the Common Neighbor Analysis (CNA) with some of its specific patterns (CNAP).
In detail, we focus on the [12,(5,5,5)] pattern that identify the presence of a Ih centre. \cite{art:sapphire}
By interpolating these values, we define the characteristic regions for different shapes.
\begin{table}[h!]
\centering
\begin{tabular}{ | p{2.5cm} | p{4cm} | p{4cm} | p{2.cm} | }
\hline
\textbf{Shape Classification} & \textbf{CNA (5,5,5) Criteria} & \textbf{Additional Criteria} & \textbf{Colormap} \\ \hline
\textbf{Decahedral (Dh)}      & CNA (5,5,5) values fall between the green and red threshold lines. & - & Blue \\ \hline
\textbf{FCC} (Face-Centered Cubic) & CNA (5,5,5) values fall below the red threshold line but above the blue threshold line. & - & Plum \\ \hline
\textbf{Icosahedral (Ih)}     & CNA (5,5,5) values lie above the green threshold dots. & At least a CNAP = [12, (5,5,5)] is present, but if they are two they do not both correspond to the seed atoms forming icosahedra. & Light Green \\ \hline
\textbf{Double-Ih}            & - & Both icosahedral centers -CNAP = [12, (5,5,5)]- are preserved from seed atoms forming icosahedra. & Dark Green \\ \hline
\end{tabular}
\caption{Classification scheme based on CNA (5,5,5) values and shape criteria.}
\end{table}
\section{Chemical Ordering classification criteria}
The chemical ordering classification is performed using the chemical Layer by Layer (cLBL) analysis method. The first step involves rotating the system so that the line connecting the centers of mass of the two chemical components aligns with the z-axis. In the second step, the nanoalloy (NA) is divided into layers, each with a thickness equal to the transition metal (TM) lattice parameter divided by $\sqrt{3}$. The chemical population of each layer is then computed. The layers are numbered starting from $z = 0$ and progressing upward, with the "1st layer" being the one closest to the transition metal side and the highest layer corresponding to the Au side.
To ensure consistent nomenclature across the analyzed systems, we standardize the classification criteria. Table \ref{tab:layer_vs_classification} summarizes the criteria for the classification. 
The cLBL results of our NAs at 200ns is shown in the upper panels figures \ref{fig:AuPd_full}, \ref{fig:AuPt_full}, \ref{fig:AuRh_full}.
\begin{table}[hp]
    \centering
    \begin{tabular}{ | c | p{2.5cm} | p{2.5cm} | p{2.5cm} | p{2.5cm} | p{2.5cm} }
    \hline
    \textbf{TM Size} & \textbf{Janus} & \textbf{S@C} & \textbf{SD} & \textbf{BC} \\
    \hline
    55   & \textit{0\% Au} in first 2 layers & \textit{Au\% in 1st layer} $>$ 75\% & \textit{Au\%} $\ne$ 100\% in last 1/4 of layers & Not matching other criteria \\
    147  & \textit{0\% Au} in first 4 layers & \textit{Au\% in 1st layer} $>$ 75\% & \textit{Au\%} $\ne$ 100\% in last 1/4 of layers & Not matching other criteria \\
    309  & \textit{0\% Au} in first 5 layers & \textit{Au\% in 1st layer} $>$ 75\% & \textit{Au\%} $\ne$ 100\% in last 1/4 of layers & Not matching other criteria \\
    561  & \textit{0\% Au} in first 6 layers & \textit{Au\% in 1st layer} $>$ 75\% & \textit{Au\%} $\ne$ 100\% in last 1/4 of layers & Not matching other criteria \\
    923  & \textit{0\% Au} in first 7 layers & \textit{Au\% in 1st layer} $>$ 75\% & \textit{Au\%} $\ne$ 100\% in last 1/4 of layers & Not matching other criteria \\
    1415 & \textit{0\% Au} in first 9 layers & \textit{Au\% in 1st layer} $>$ 75\% & \textit{Au\%} $\ne$ 100\% in last 1/4 of layers & Not matching other criteria \\
    \hline
    \end{tabular}
    \caption{Classification criteria for systems based on cLBL. The table outlines the specific conditions used to categorize each system according to the chemical composition of specific layers. }
    \label{tab:layer_vs_classification}
\end{table}

Along with the chemical LBL distribution we considered also the most used chemical radial distribution function (cRDF).
The radial distribution function in a system of particles, describes how density varies as a function of distance from a reference particle. It is a measure of the probability of finding a particle at a distance of r away from a given reference particle, here the center of mass.
Calculating the density of atoms from the centre of mass of the whole nanoparticle ($\mathrm{CoM}_w$); the distribution of atoms-A from their centre of mass, $\mathrm{CoM}_A$; and similarly for B-atoms, from the $\mathrm{CoM}_B$.
The RDFs count the number of atoms falling in concentric shells from the centre of mass, of the nanoparticle:
\begin{equation}
r_{\alpha}(i \in w,A,B)=\sqrt{\left(\hat{x}_\alpha(i)\right)^2+\left(\hat{y}_\alpha(i)\right)^2+\left(\hat{z}_\alpha(i)\right)^2} .
\label{eq:radRDF}
\end{equation}
where the coordinates $w$ is referring to whole cluster, $\hat{x}_\alpha, \hat{y}_\alpha, \hat{z}_\alpha$ of the atom-i and chemical species $\alpha=A, B$ are re-scaled w.r.t the centre of mass chosen, $\mathrm{CoM}_w$, $\mathrm{CoM}_A$ or $\mathrm{CoM}_B$.
In detail, we defined a radial distribution funcion of whole cluster RDF and a chemical radial distribution function CRDF that consider the radial distribution function for the chemical element choosen.
This descriptor can give us quantitative information about the distribution of atoms, but in our case, considering the asphericity of our NAs, it is not trivial to use it for categorizing uniquely in one of the well-known chemical orderings (mixed, ball-cup, core-shell, off-centered core-shell). 
In our case, the cLBL distribution results to be more effective for a classification, also if we consider the cRDF for a double check over the results. The results at 200 ns for the cRDF are provided in the lower panels of Figures \ref{fig:AuPd_full}, \ref{fig:AuPt_full}, \ref{fig:AuRh_full}.

\noindent\includegraphics[width=1\linewidth]{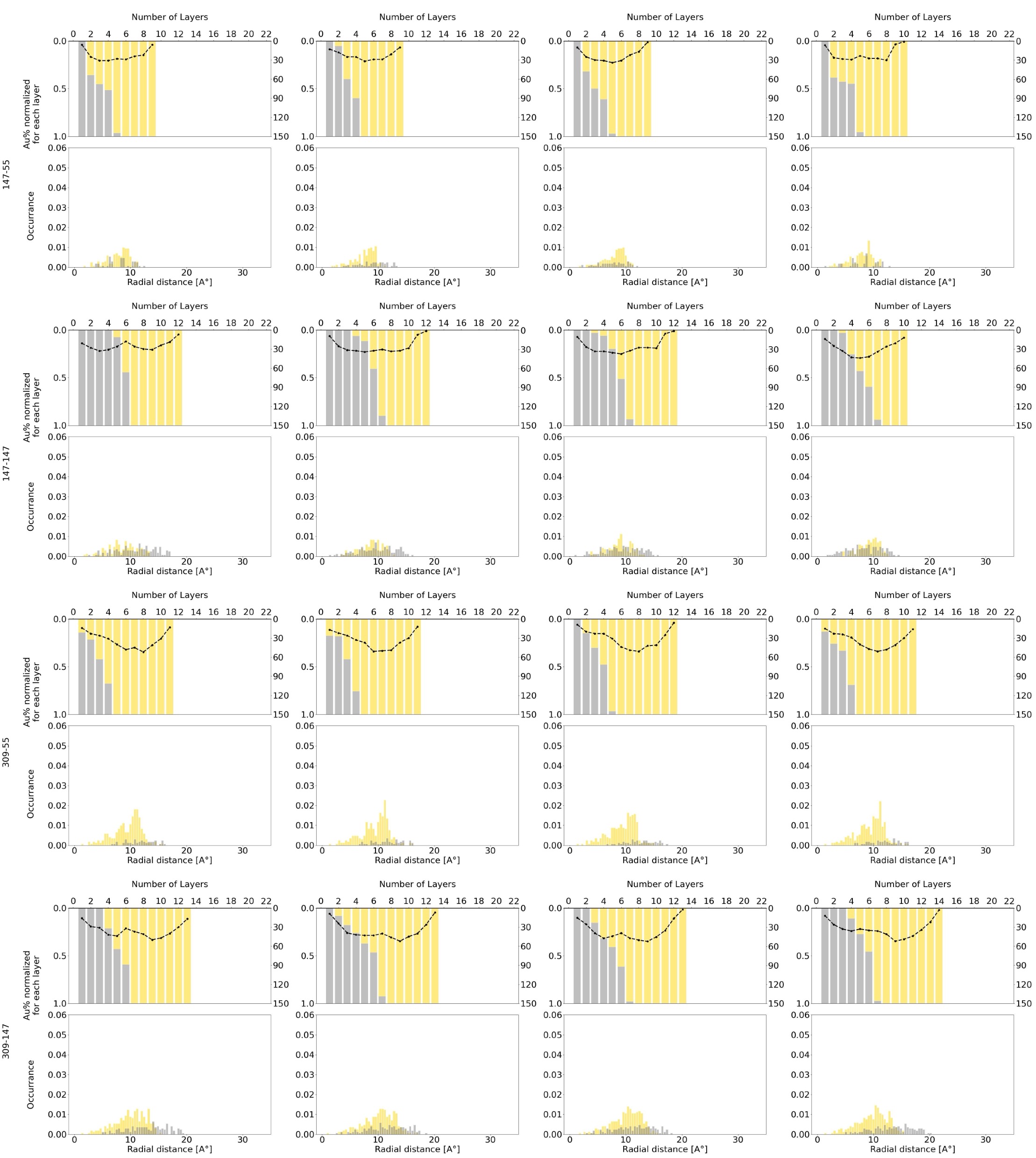}

\vspace{0.3cm}

\noindent\includegraphics[width=1\linewidth]{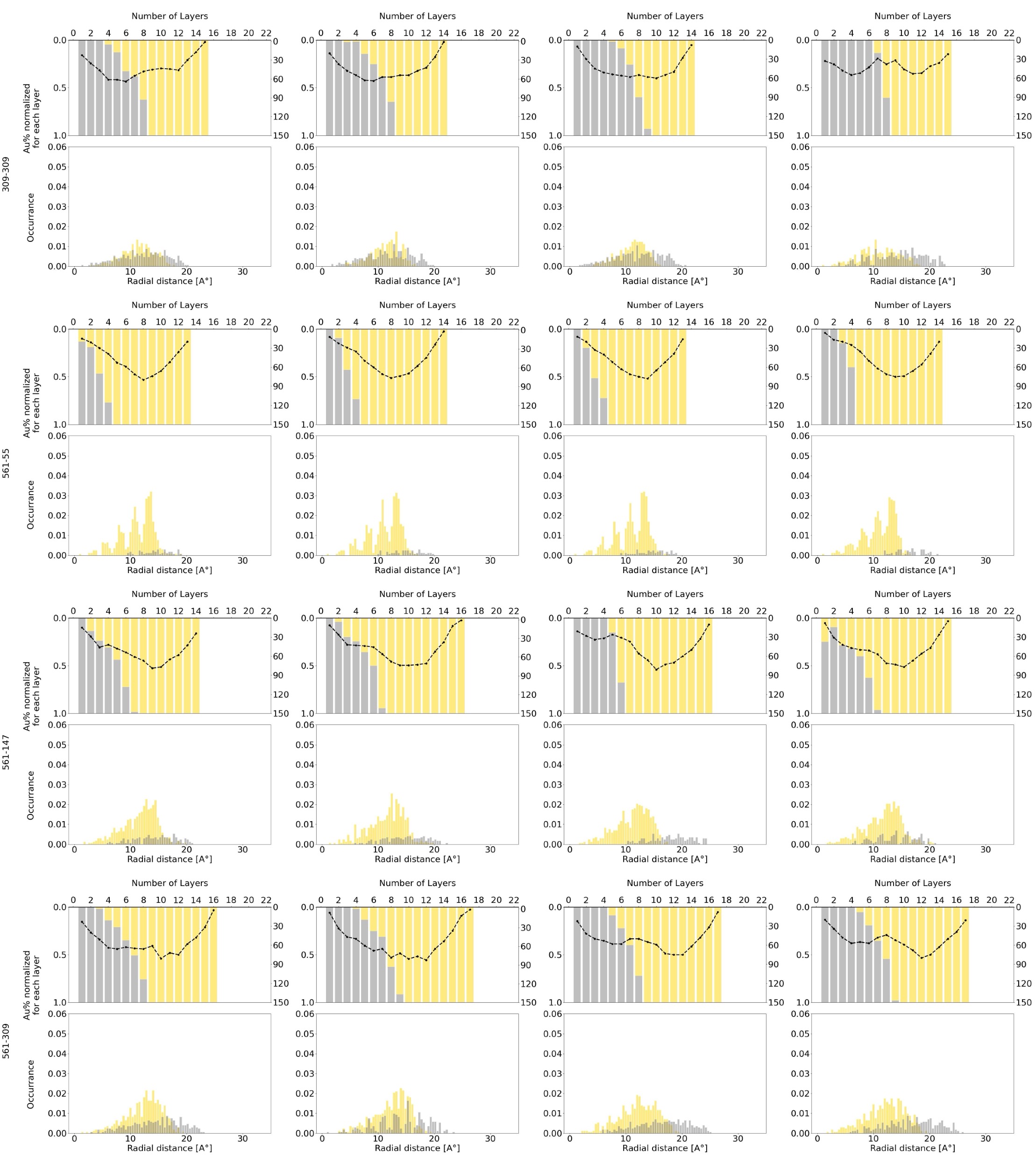}

\vspace{0.3cm}

\noindent\includegraphics[width=1\linewidth]{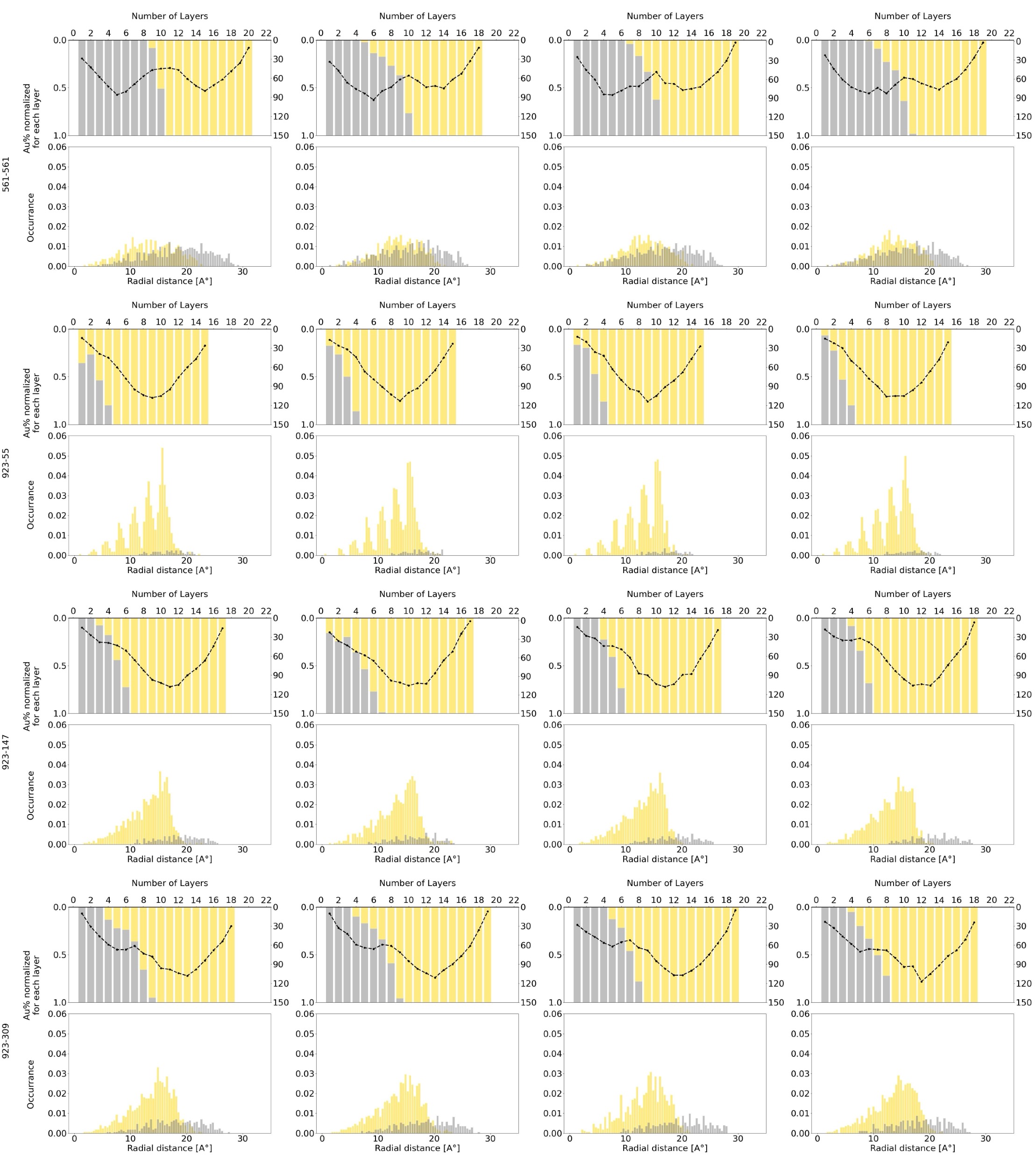}

\vspace{0.3cm}

\noindent\includegraphics[width=1\linewidth]{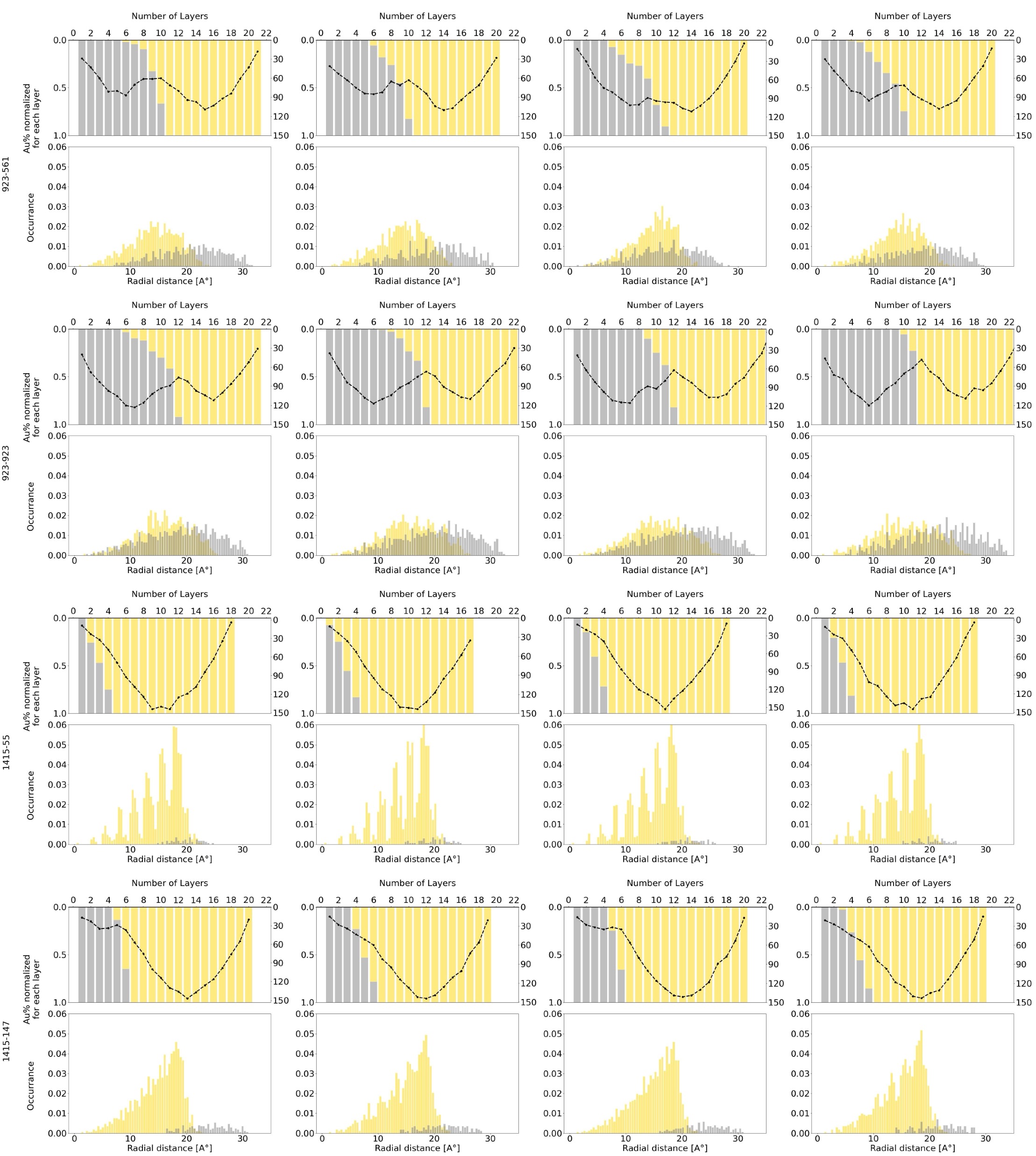}

\vspace{0.3cm}

\noindent\includegraphics[width=1\linewidth]{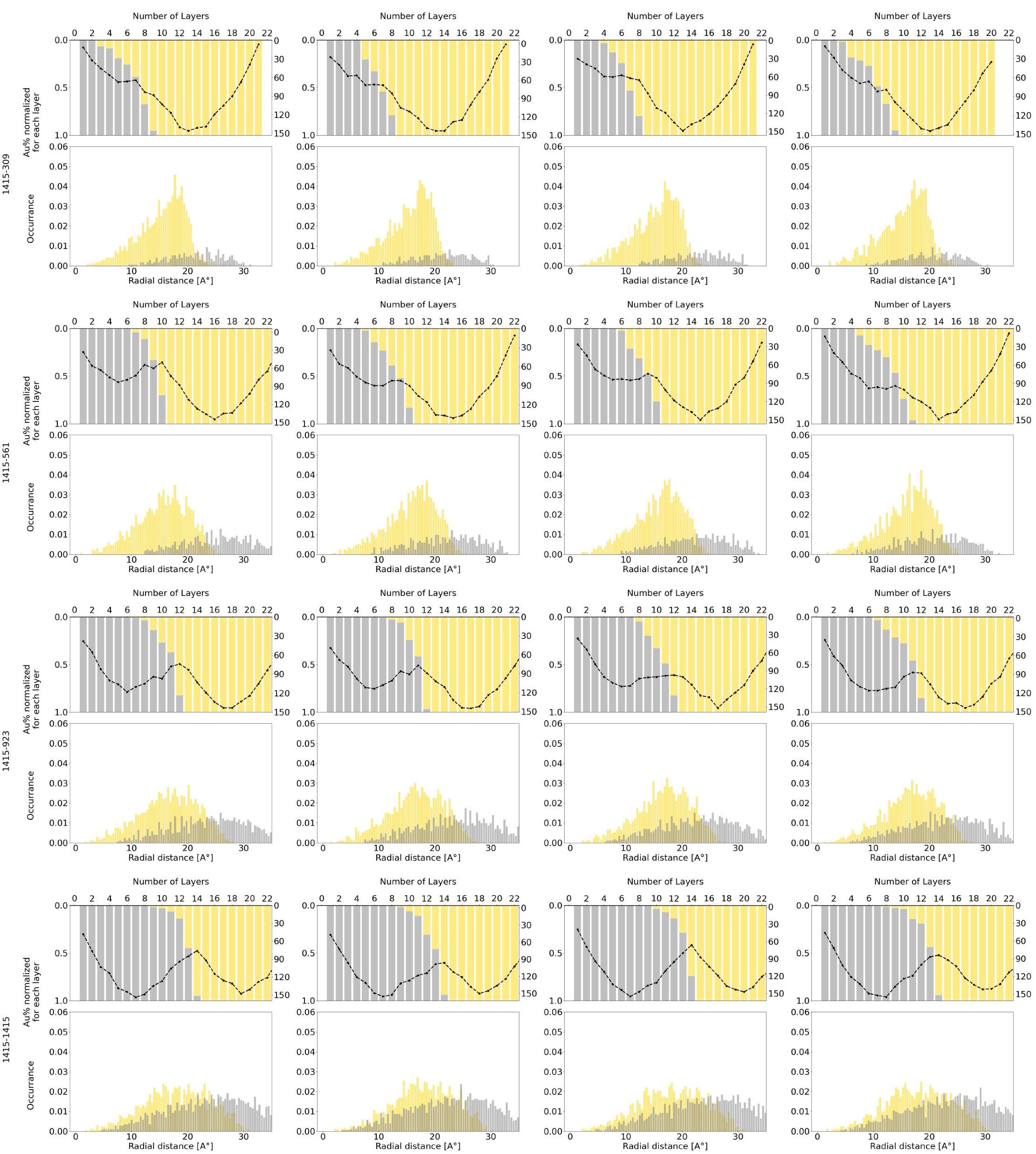}

\captionof{figure}
{AuRh. In the upper panel, LBL analysis: black dotted line, and cLBL histogram reporting the percentage of TM in grey and gold in yellow, normalized to the total number of atoms for each layer. The minimum of the LBL corresponds to the most populated layer. In the lower panel, the cRDF: TM in grey, and Au in yellow.}
\label{fig:AuRh_full}

\vspace{1cm}

\noindent\includegraphics[width=1\linewidth]{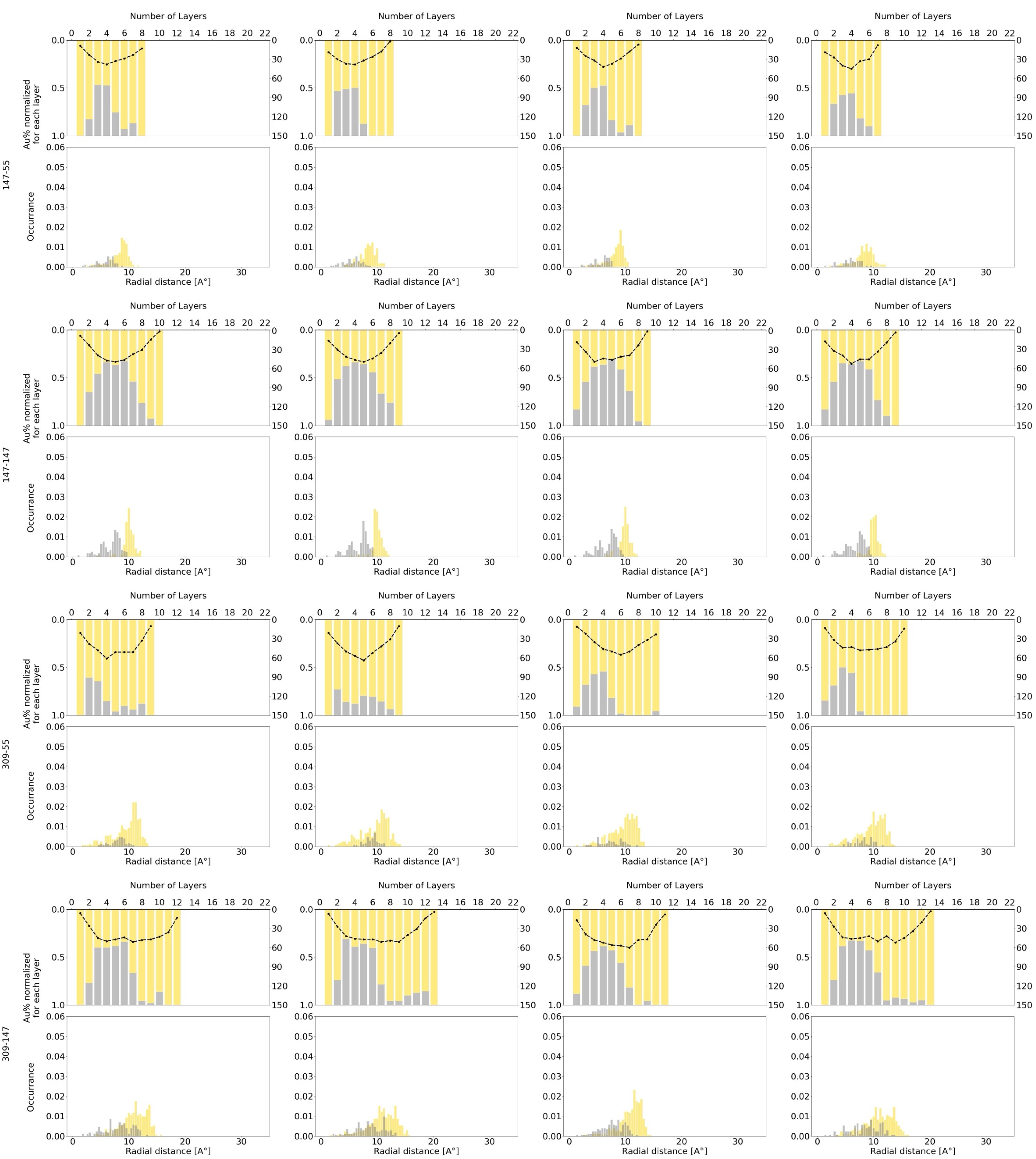}

\vspace{0.3cm}

\noindent\includegraphics[width=1\linewidth]{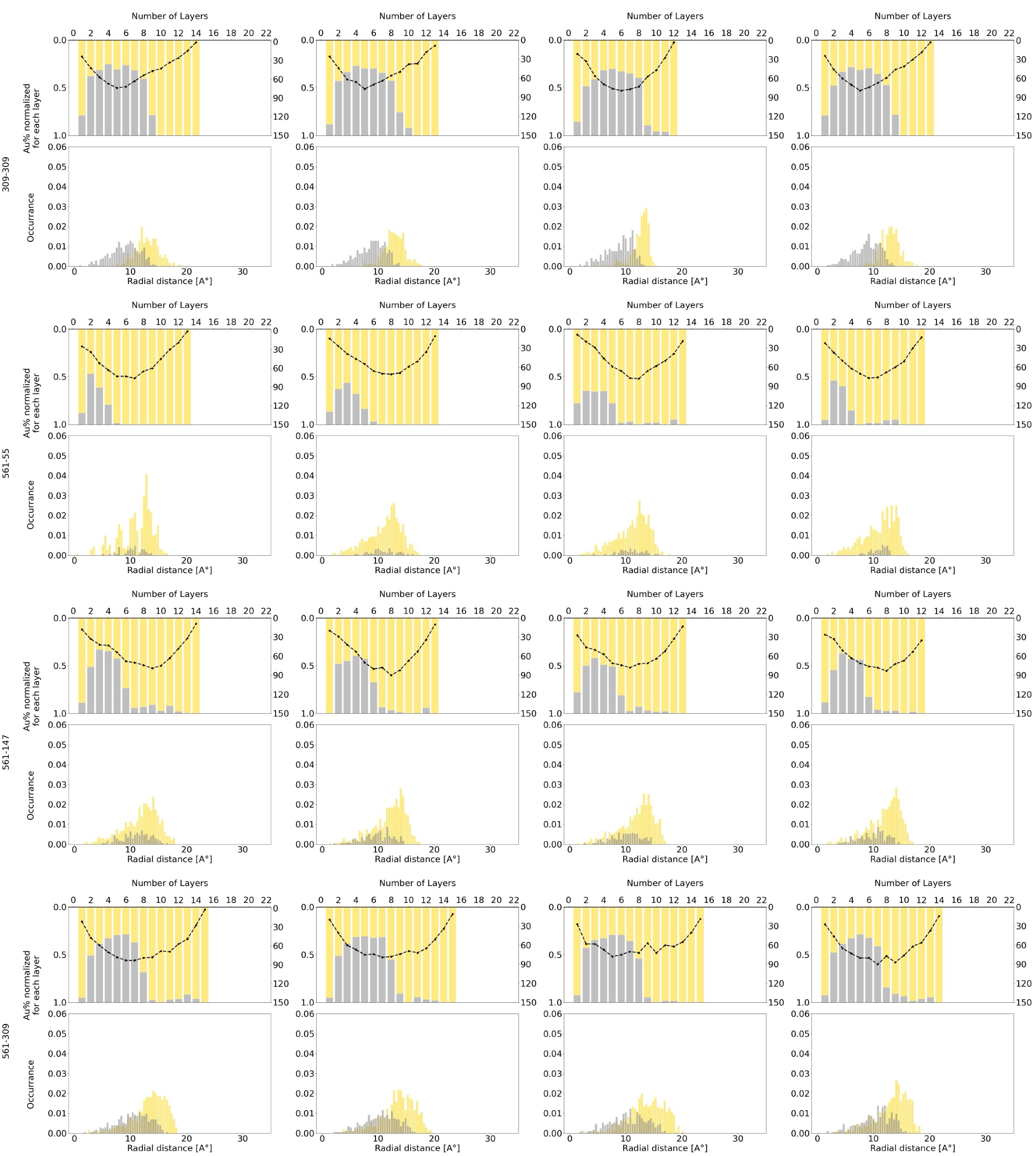}

\vspace{0.3cm}

\noindent\includegraphics[width=1\linewidth]{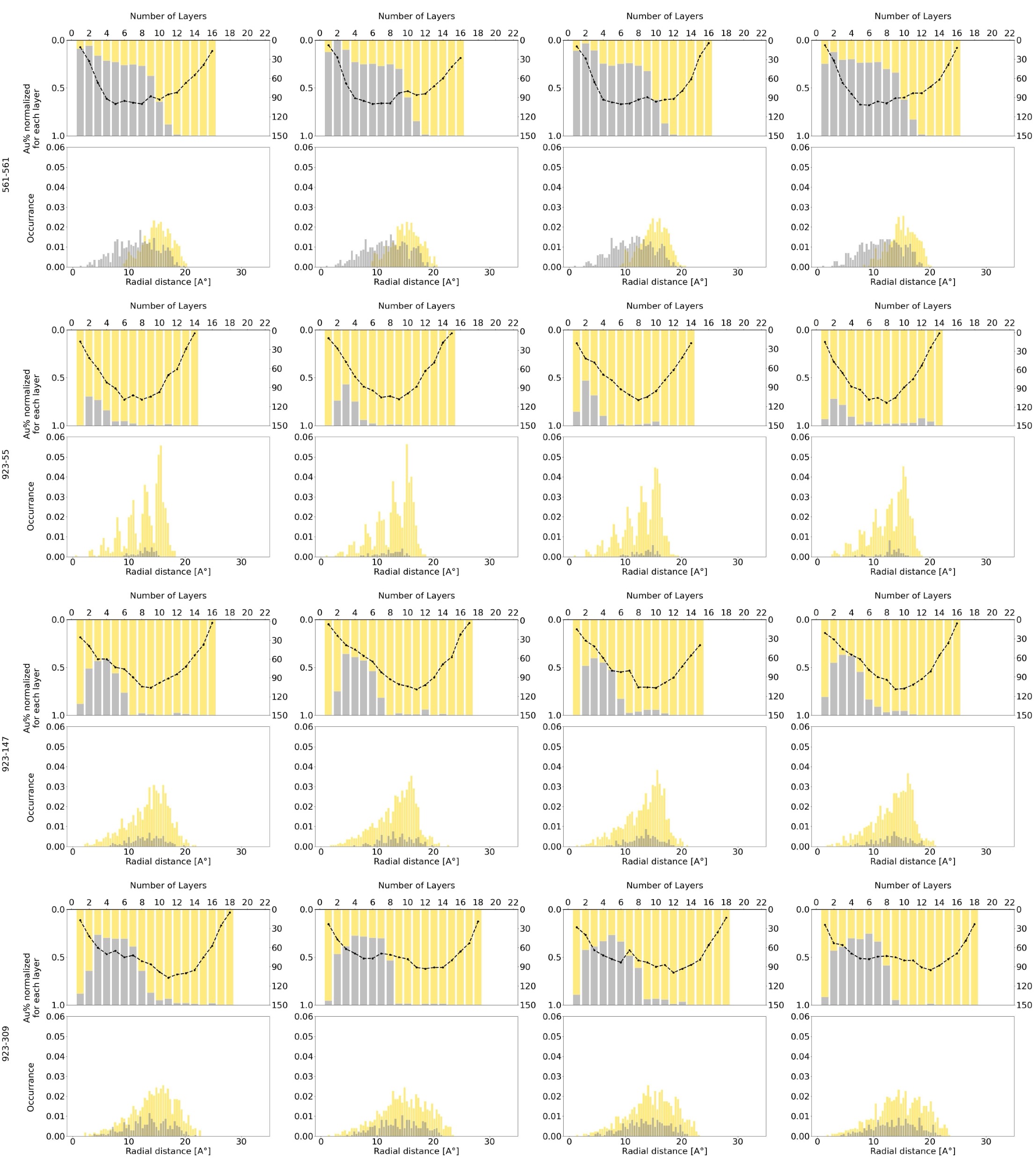}

\vspace{0.3cm}

\noindent\includegraphics[width=1\linewidth]{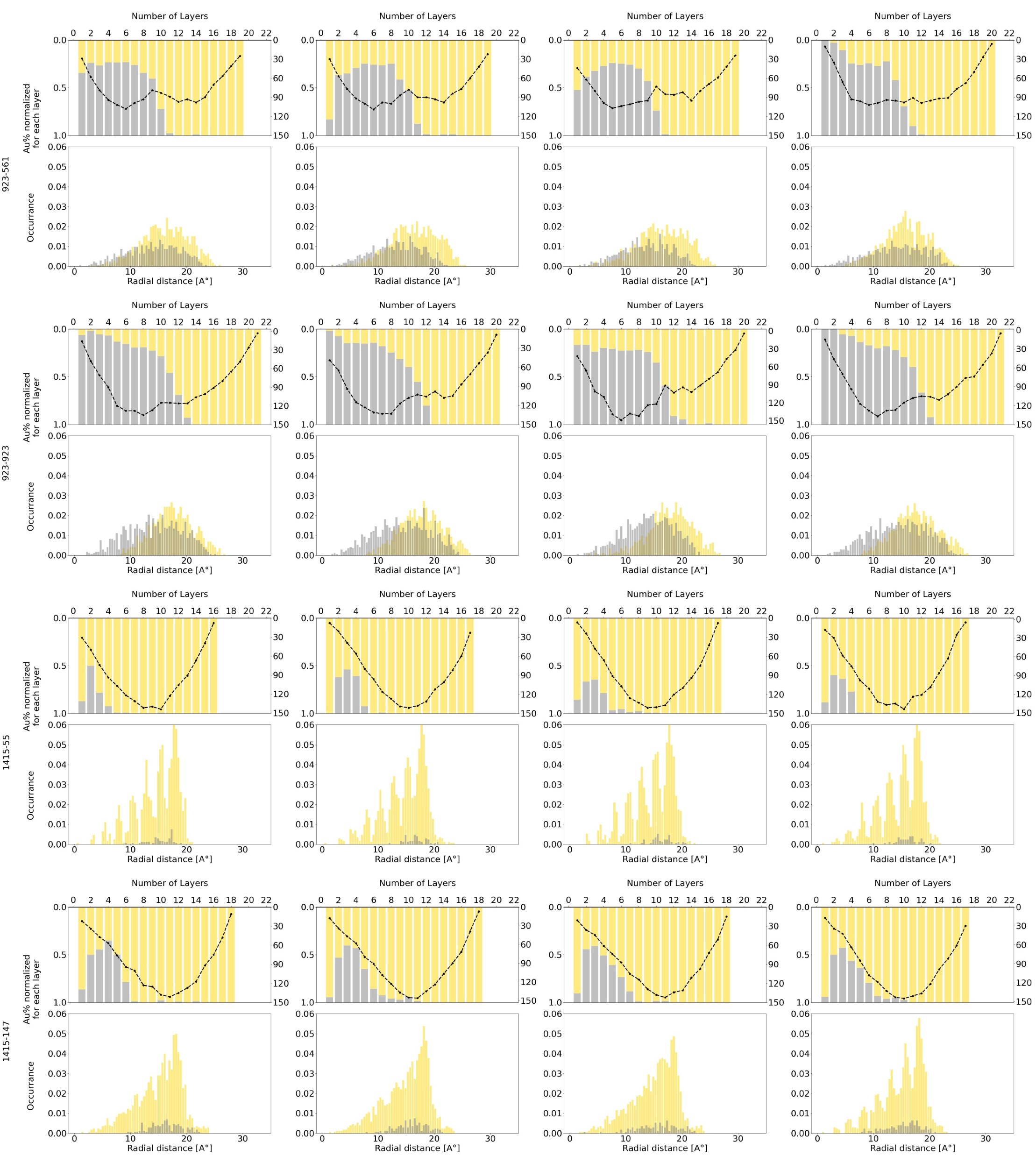}

\vspace{0.3cm}

\noindent\includegraphics[width=1\linewidth]{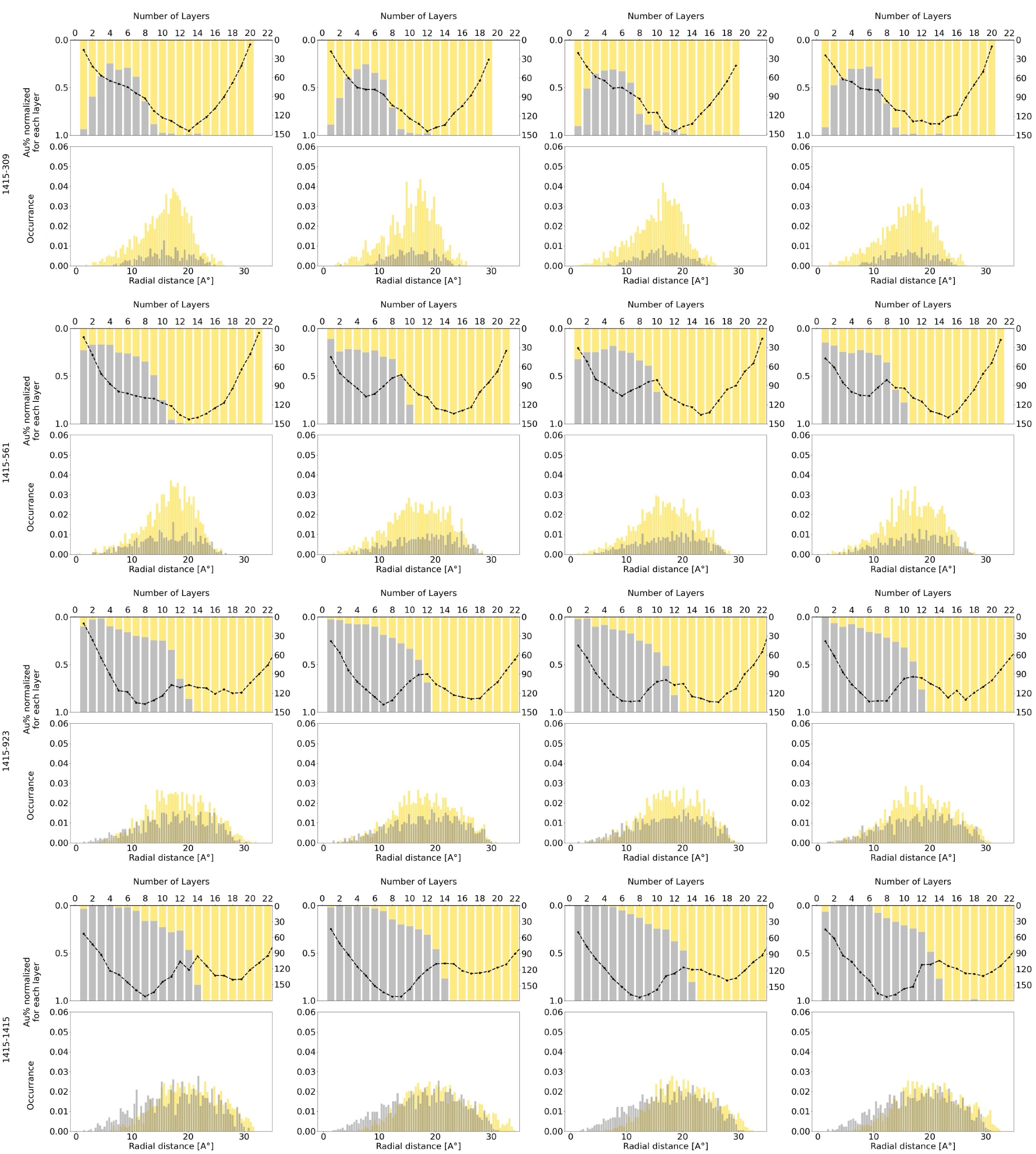}

\captionof{figure}{ 
AuPt. In the upper panel, LBL analysis: black dotted line, and cLBL histogram reporting the percentage of TM in grey and gold in yellow, normalized to the total number of atoms for each layer. The minimum of the LBL corresponds to the most populated layer. In the lower panel, the cRDF: TM in grey, and Au in yellow.}
\label{fig:AuPt_full}

\noindent\includegraphics[width=1\linewidth]{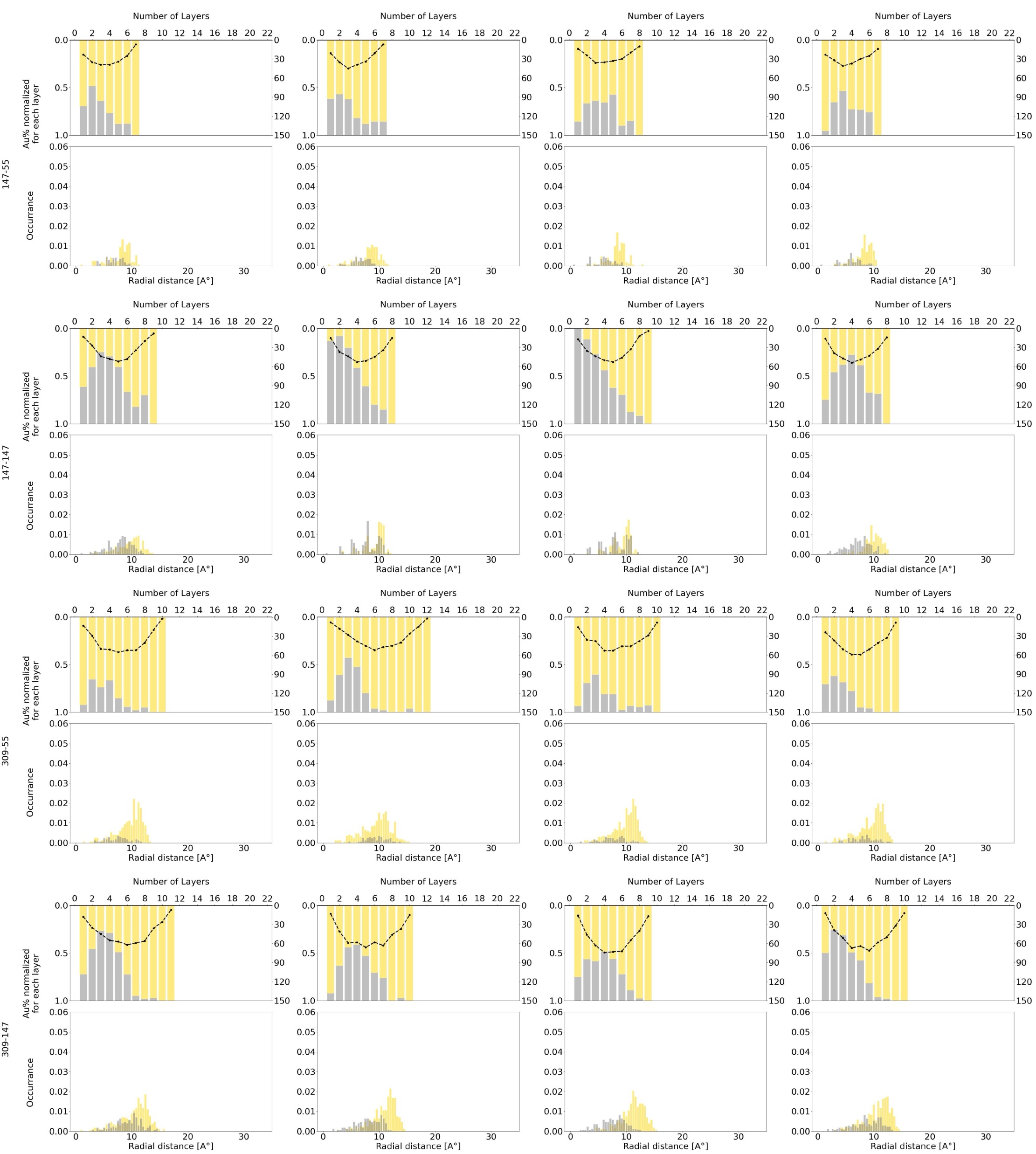}

\vspace{0.3cm}

\noindent\includegraphics[width=1\linewidth]{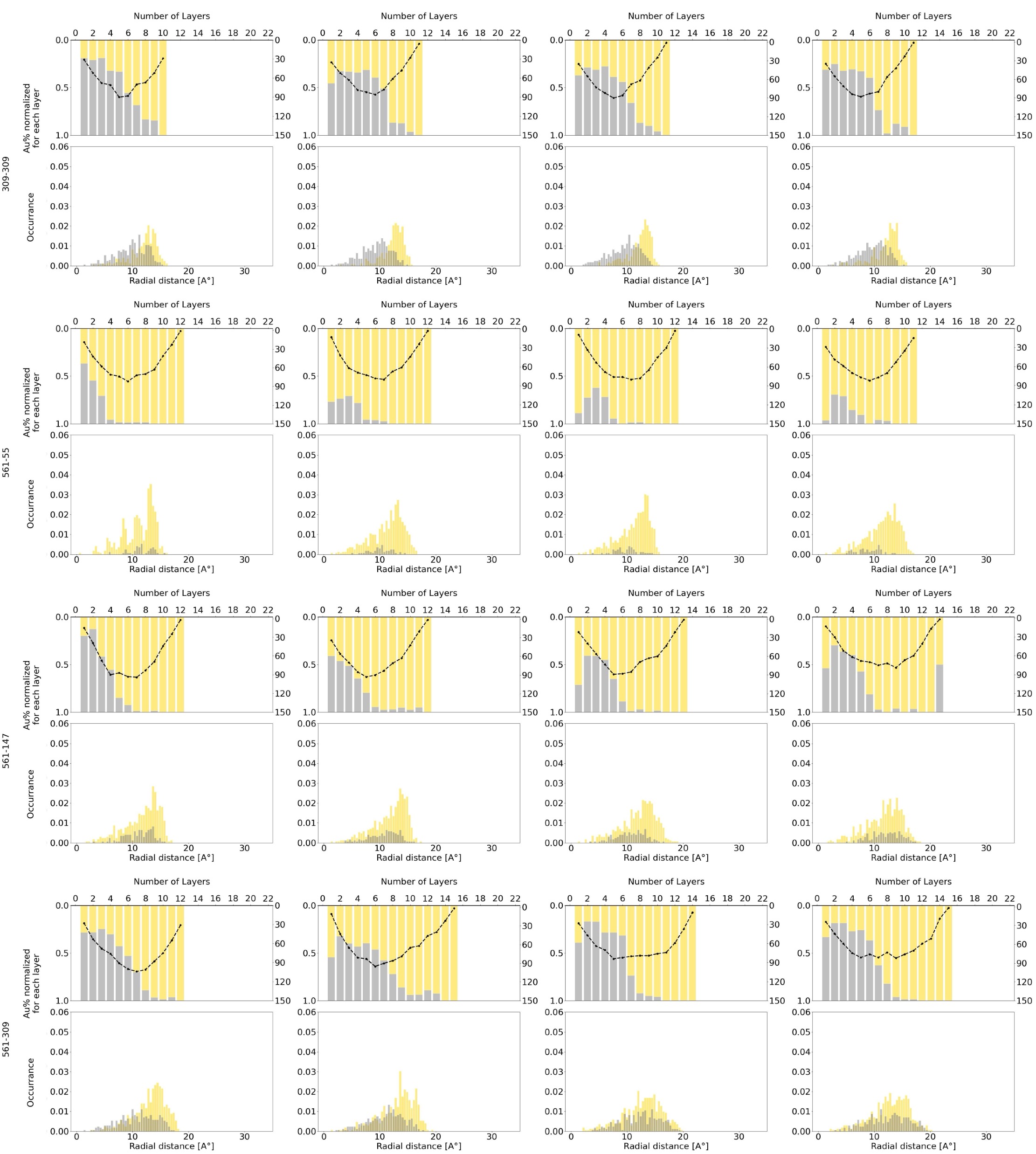}

\vspace{0.3cm}

\noindent\includegraphics[width=1\linewidth]{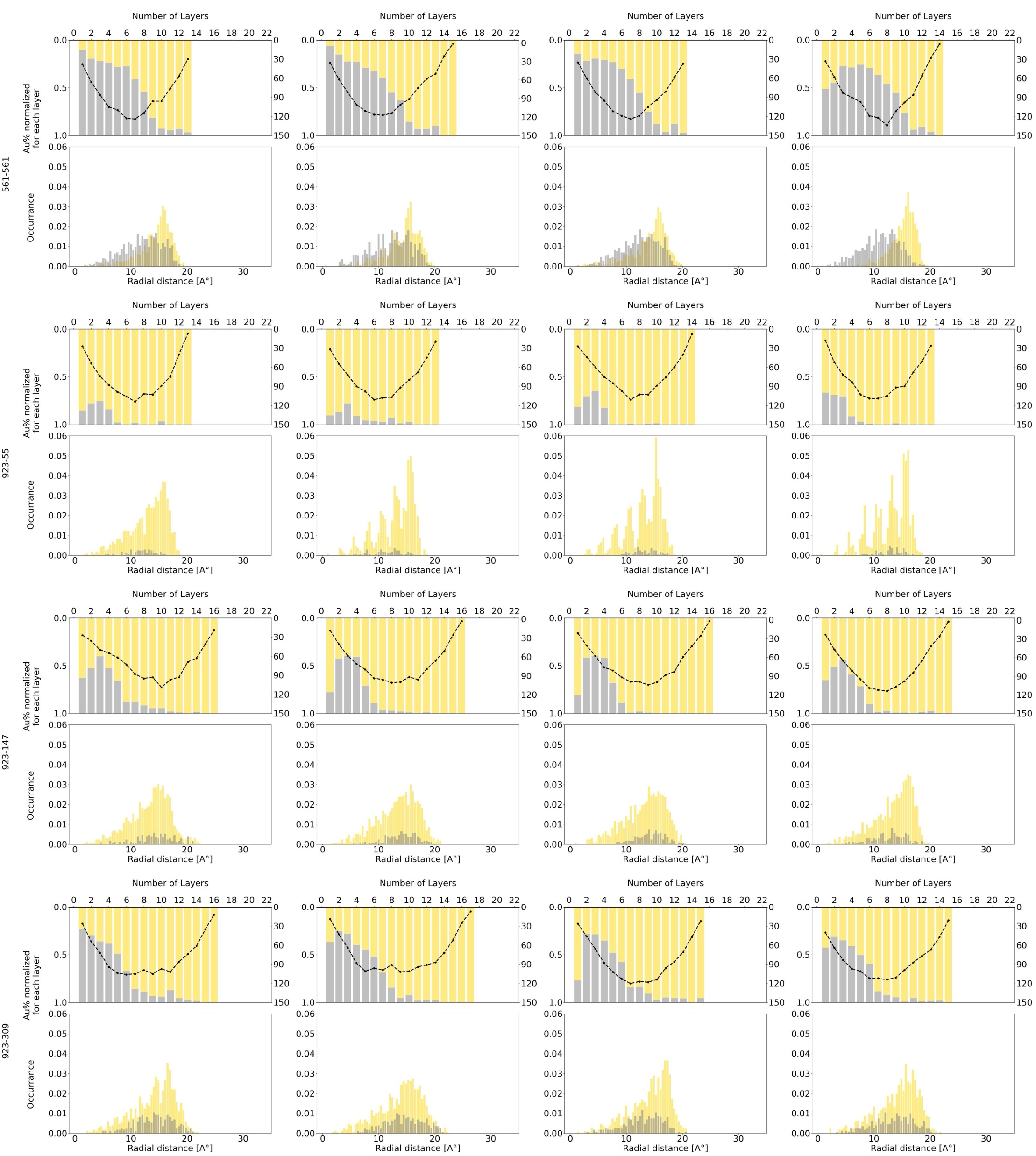}

\vspace{0.3cm}

\noindent\includegraphics[width=1\linewidth]{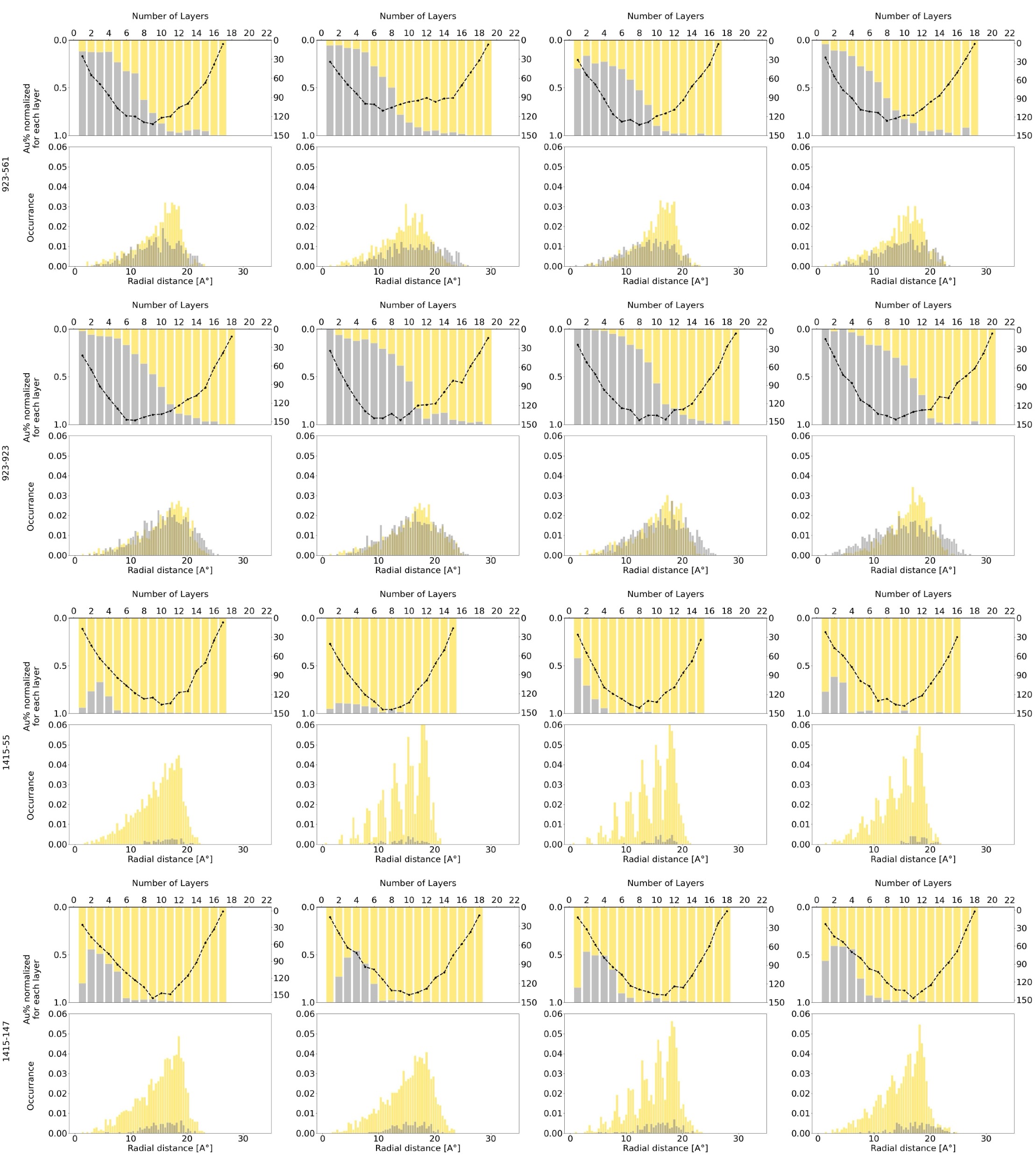}

\vspace{0.3cm}

\noindent\includegraphics[width=1\linewidth]{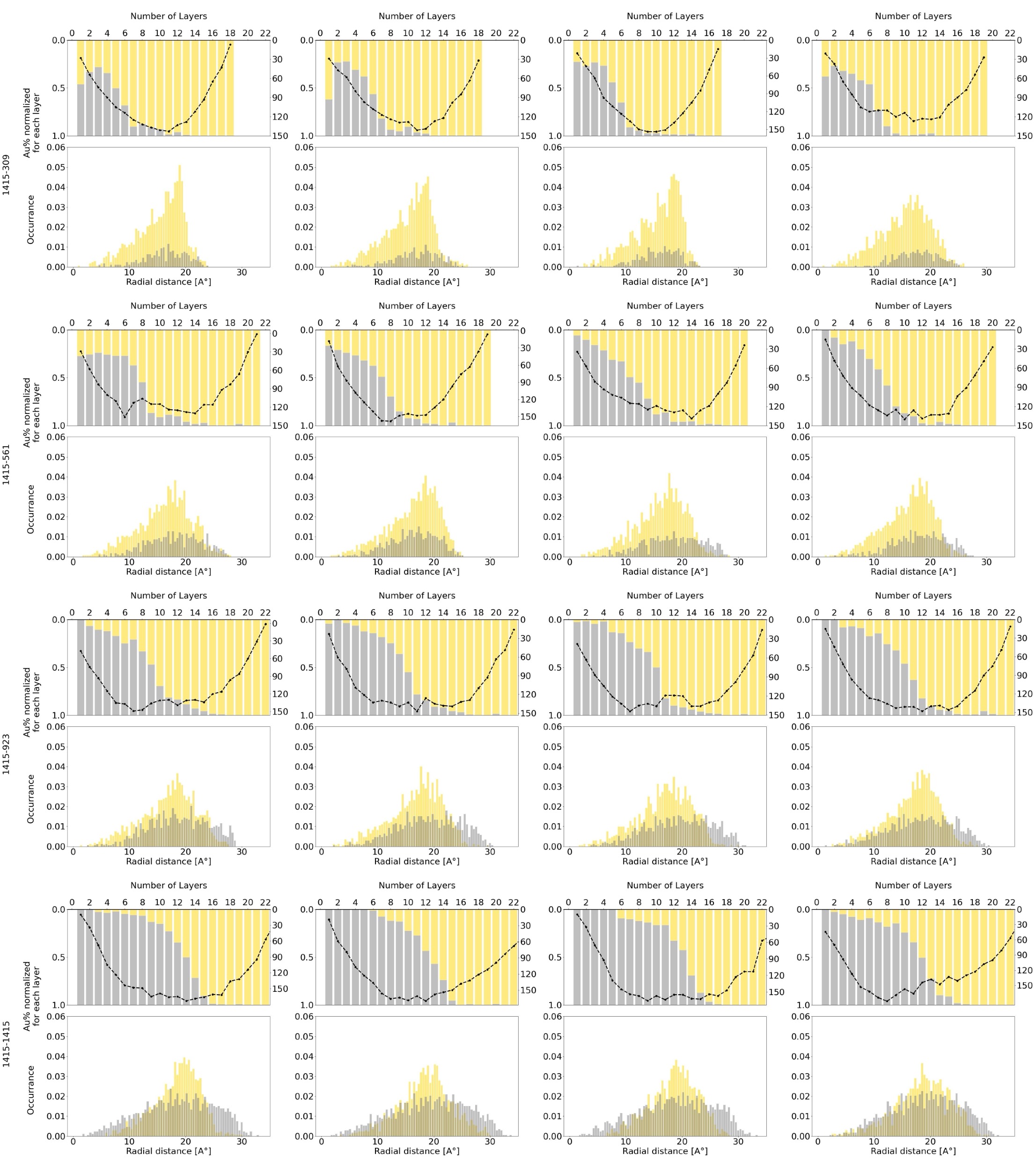}

\captionof{figure}{
AuPd. In the upper panel, LBL analysis: black dotted line, and cLBL histogram reporting the percentage of TM in grey, and gold in yellow, normalized to the total number of atoms for each layer. The minimum of the LBL corresponds to the most populated layer. In the lower panel the cRDF: TM in grey, and Au in yellow.}
\label{fig:AuPd_full}

\vspace{1cm}

\section{Local Atomic Environment groups and Mean Square Displacement}
The Local Atomic Environment (LAE) \cite{art:sapphire} quantifies the chemical population -nearest neighbours (NN)- within the first coordination shell, corresponding to the first minimum of the pair-distance distribution function (PDDF). 
The classification into four groups can be expressed as:

$$
\mbox{
LAE = }
\begin{cases}
  \mbox{LAE$_1$}, & \text{if } NN_{\text{other species}, i} = 0  \\
  \mbox{LAE$_2$}, & \text{if } 1 \leq NN_{\text{other species}, i} \leq 2 \\
  \mbox{LAE$_3$}, & \text{if } 3 \leq NN_{\text{other species}, i} \leq 8 \quad \mbox{(interface)} \\
  \mbox{LAE$_4$}, & \text{if } NN_{\text{other species}, i} \geq 9 \quad \mbox{(atom dispersed in opposite element matrix)} \\
\end{cases}
$$

Each atom is part of one of these four groups and for simplicity we will refers to them as group LAE$_1$, LAE$_2$, LAE$_3$, LAE$_4$ if we are referring to Au and LAE$_5$, LAE$_6$, LAE$_7$, LAE$_8$ if we are referring to the TM, respectively.

%
%

\begin{sidewaysfigure}[htbp]
\includegraphics[width=.9\linewidth]{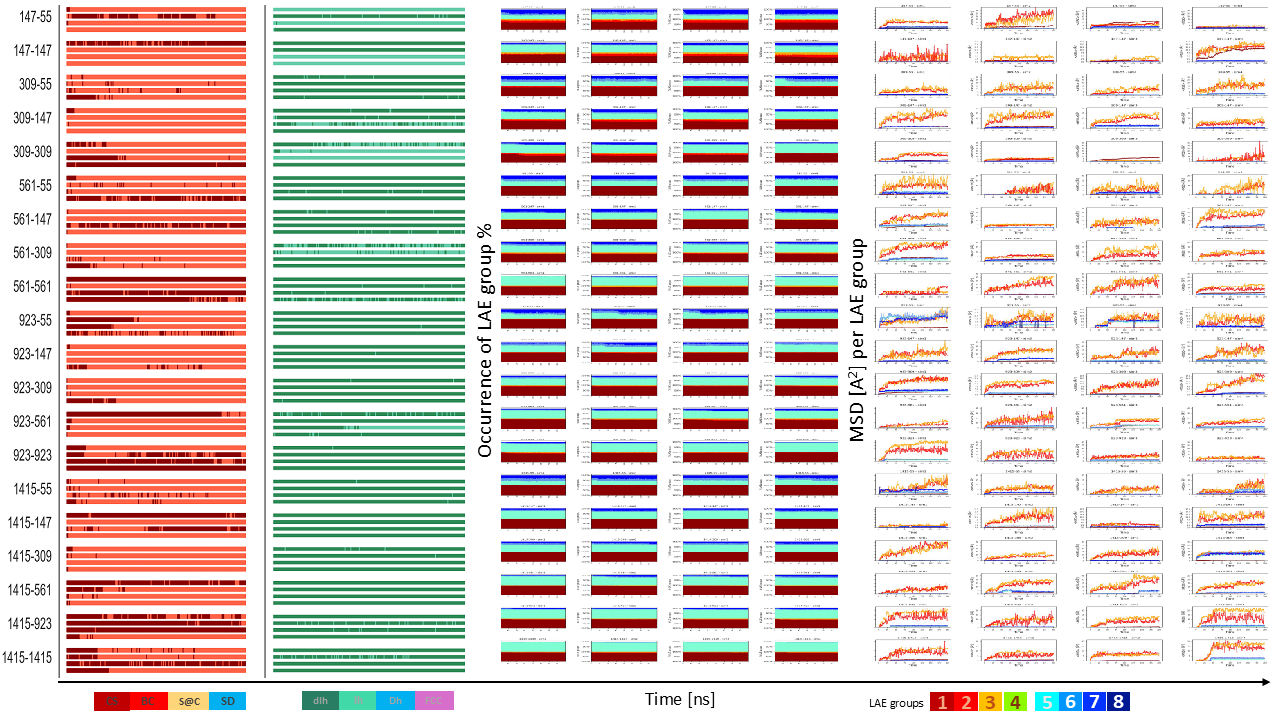}
    \centering
    \caption{Comprehensive analysis of AuRh core-satellites nanoparticles. From left to right:
    (i) Chemical ordering evolution; 
    (ii) Structural evolution via CNA; 
    (iii) Temporal evolution of LAE populations; 
    (iv) MSD of atoms in each LAE group.
    Rows are organized from top to bottom by increasing Au-core size (number of Au atoms indicated at the beginning of each row). Colors assigned according to the legend provided in the text.}
    \label{fig:TM_AuRh}
\end{sidewaysfigure}

\begin{sidewaysfigure}[htbp]
\includegraphics[width=.9\linewidth]{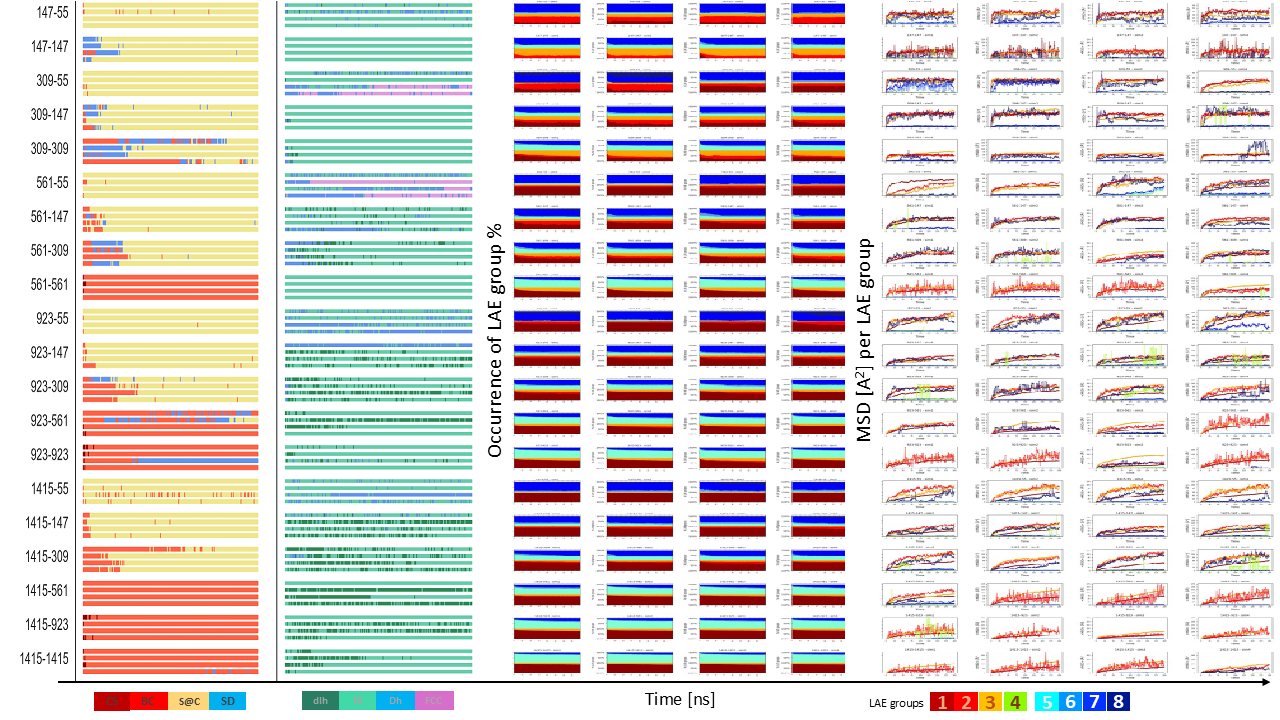}
    \centering
    \caption{Comprehensive analysis of AuPt core-satellites nanoparticles. From left to right:
    (i) Chemical ordering evolution; 
    (ii) Structural evolution via CNA; 
    (iii) Temporal evolution of LAE populations; 
    (iv) MSD of atoms in each LAE group.
    Rows are organized from top to bottom by increasing Au-core size (number of Au atoms indicated at the beginning of each row). Colors assigned according to the legend provided in the text.}
    \label{fig:TM_AuPt}
\end{sidewaysfigure}

\begin{sidewaysfigure}[htbp]
\includegraphics[width=.9\linewidth]{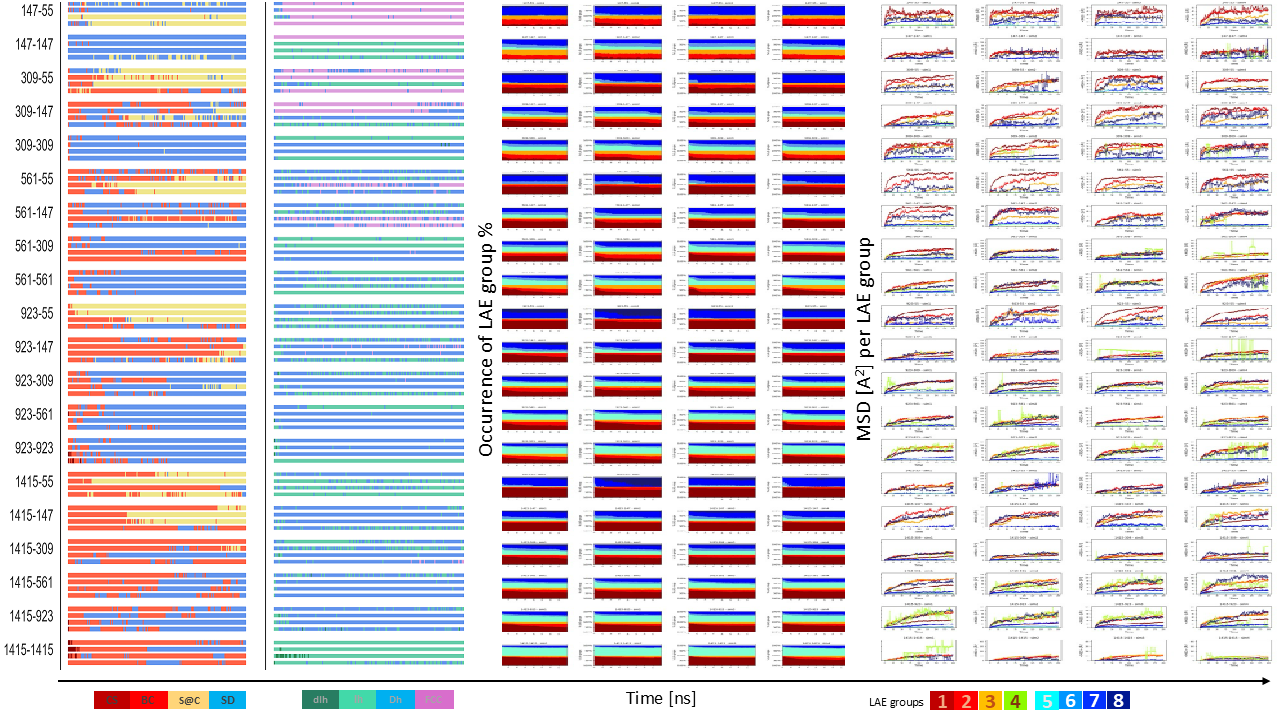}
    \centering
    \caption{Comprehensive analysis of AuPd core-satellites nanoparticles. From left to right:
    (i) Chemical ordering evolution; 
    (ii) Structural evolution via CNA; 
    (iii) Temporal evolution of LAE populations; 
    (iv) MSD of atoms in each LAE group.
    Rows are organized from top to bottom by increasing Au-core size (number of Au atoms indicated at the beginning of each row). Colors assigned according to the legend provided in the text.}
    \label{fig:TM_AuPd}
\end{sidewaysfigure}

\begin{figure}[htbp]
    \centering
    \includegraphics[width=.9\linewidth]{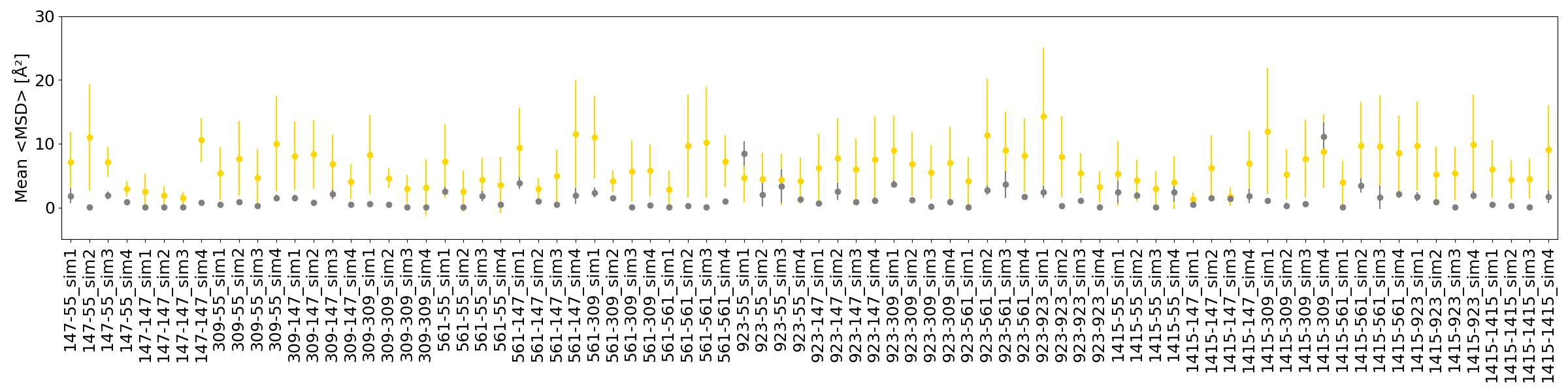}
    \centering
    \includegraphics[width=.9\linewidth]{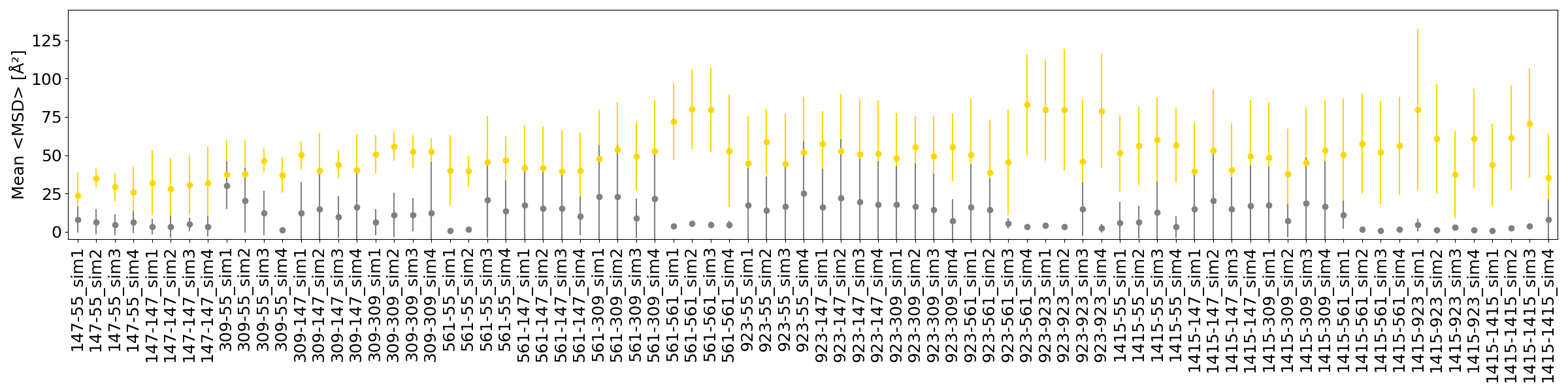}
    \centering
    \includegraphics[width=.9\linewidth]{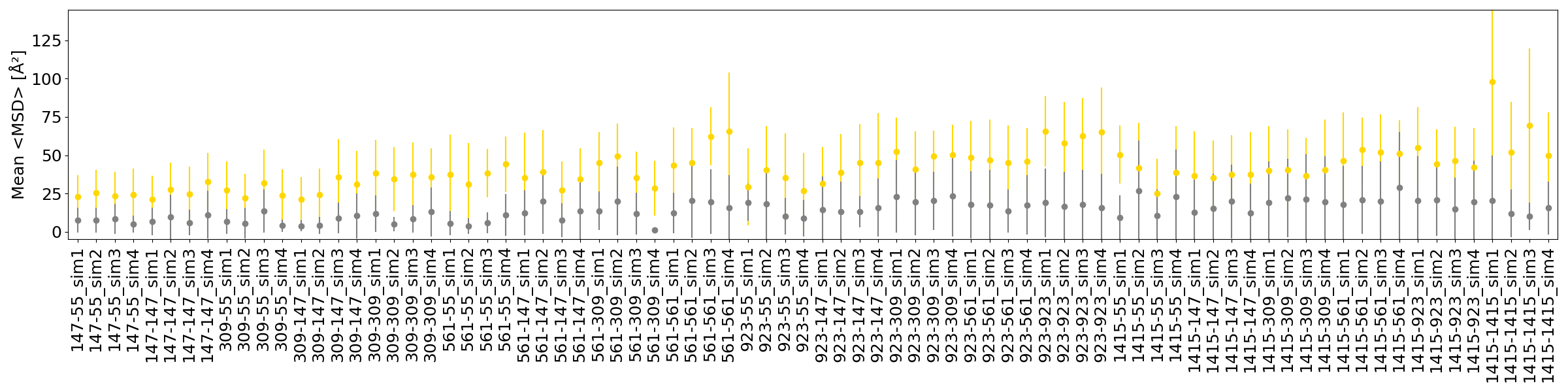}
    \caption{
    Averaged MSD analysis of - from top to bottom respectively - AuRh, AuPt, and AuPd  core-satellites nanoparticles.
    MSD mean value averaging over time and over the four Au LAE-groups and the four TM LAE-groups, in yellow and gray respectively.
    }
    \label{fig:MSD_medie}
\end{figure}

We also applied the calculation of the Mean Square Displacement (MSD) with respect to the structure at $t_0$=1ns, as:
$$
\mbox{MSD} (t) = \langle |\mathbf{r}(t) - \mathbf{r}(t_0)|^2 \rangle
$$
Where:
$\mathbf{r}(t)$ is the position of the atom at time $t$,
$\mathbf{r}(t_0)$ is the position of the atom at time $t_0 = 1 \, \text{ns}$, $\langle \dots \rangle$ represents an ensemble average over all atoms.
This equation computes the average squared displacement of atoms from their initial positions at time $t_0$, giving an indication of how far atoms have moved over time.
In Figures \ref{fig:TM_AuRh}-\ref{fig:TM_AuPt}-\ref{fig:TM_AuPd} we report the complete results for the percentage of atoms in each LAE group for each size and independent simulation and the mean MSD for each LAE group, while in \ref{fig:MSD_medie} we show the mean value of the MSD that each simulation possess, for Au LAE-groups and TM LAE-groups.
In detail, in Figures \ref{fig:TM_AuRh}-\ref{fig:TM_AuPt}-\ref{fig:TM_AuPd} from left to right: (i) Chemical ordering evolution of Au-TM core–satellite nanoparticles. Each row corresponds to an independent simulation. Structures are sampled every 1 ns. Each point denotes the classification of chemical ordering: Janus (dark red), ball-cup (BC, light red), Shell-Core (S@C, khaki), and semi-dispersed (SD, light blue).
(ii) Structural evolution analyzed via Common Neighbor Analysis (CNA). Identified motifs include: double icosahedral (dIh, dark green), icosahedral (Ih, light green), decahedral (Dh, blue), and face-centered cubic (FCC, plum). Each row corresponds to an independent simulation. 
(iii) Temporal evolution of local atomic environment (LAE) group populations. Each column corresponds to an independent simulation. Colors go from dark red, red, orange, gold, lime, aquamarine, royal blue, dark blue, and corresponds to LAE$_1$, LAE$_2$, LAE$_3$, LAE$_4$, LAE$_5$, LAE$_6$, LAE$_7$, LAE$_8$ respectively. 
(iv) Mean squared displacement (MSD) analysis of atoms within each LAE group. Same group colors as in the previous panel.


\vspace{0.3cm}

\vspace{0.3cm}

\clearpage

\bibliography{acs-bibliography}